\documentclass[twocolumn,amsmath,amssymb,
aps,showpacs,graphicx]{revtex4-2}

\usepackage{graphicx}
\usepackage{dcolumn}
\usepackage{bm}
\usepackage{lineno, hyperref}
\usepackage{natbib}
\begin{document}

\title{Applied electric and magnetic field effects on the bandgap formation and antiferromagnetic ordering in AA-stacked Bilayer Graphene}
\author{V. Apinyan} 
\altaffiliation[e-mail:]{v.apinyan@intibs.pl}
\author{T. Kope\'c} 
\affiliation{Institute of Low Temperature and Structure Research, Polish Academy of Sciences\\ 50-422, ul. Ok{\'o}lna 2, Wroc\l{}aw, Poland 
}
%
\begin{abstract}
In this study, we consider a two-layer graphene structure stacked in the AA form and exposed to the influence of two different electric fields applied to different layers. The graphene layers are also subjected to an external magnetic field perpendicular to the planes of the layers. We investigate the possible effects of the applied in-plane fields and the magnetic field on excitonic pairing, antiferromagnetic order, and the chemical potential. Simultaneously, we analyze the effects of the interlayer Coulomb interaction potential on the physical properties of the considered system. We demonstrate that the application of planar electric fields leads to the formation of an unusually large bandgap in the electronic band structure, which is not typical for AA-stacked bilayer graphene. We discuss various values of the applied electric field potentials and show their influence on the electronic band structure of the system. Additionally, we identify the existence of a critical value of the magnetic field above which Wigner crystallization-like effect is present for the electrons, also affecting the excitonic gap in one spin channel. The results obtained in this study could be important for applications of AA-stacked bilayer graphene as a large band-gap material.
\end{abstract}     



\maketitle
\section{\label{sec:Section_1} Introduction}

The electronic correlations in monolayer and bilayer graphene (BLG) structures represent a remarkable field of research in condensed matter physics \cite{cite_1, cite_2, cite_3, cite_4}. The unusual electronic properties of graphene provide new insights into its applications in electronics \cite{cite_5, cite_6, cite_7}.

Gating bilayer graphene and applying a magnetic field lead to intriguing physical phenomena, as the electrons change their transport behavior \cite{cite_8, cite_9, cite_10, cite_11, cite_12, cite_13, cite_14}. Graphene and graphene-based materials are also noteworthy for their excitonic, optical, and conductivity properties \cite{cite_15, cite_16, cite_17, cite_18, cite_19, cite_20, cite_21, cite_22, cite_23, cite_24, cite_25, cite_26, cite_27, cite_28}. The possibility of forming excitons in AA-stacked BLG structures has been discussed in several works. Excitons, which are coupled formations of electrons and holes, could facilitate entire phase transition regions, leading to the so-called excitonic insulator phase transition in graphene systems \cite{cite_29, cite_30, cite_31,cite_32,cite_33}. The coexistence of the excitonic phase with the antiferromagnetic phase, under the influence of applied electric field potentials, has been demonstrated in a series of studies \cite{cite_34, cite_35, cite_36, cite_37, cite_38}.

References \cite{cite_34, cite_37} indicate that the antiferromagnetic order parameter is significantly larger than the excitonic one in AA-type BLG. Conversely, Ref. \cite{cite_38} reports that in the AB Bernal BLG system, the excitonic order parameter is dominant.

The existence of the Wigner crystal phases has been shown in bilayer graphene system, in the context of the screening effects of local the Coulomb interaction \cite{cite_39} and it has been shown that in-plane charge inhomogeneity is impossible in the undoped bilayer graphene. Moreover, it has been shown that the magnetic field stabilizes the Wigner crystal phases in graphene moir\'e superlattices \cite{cite_40}.    

In Ref. \cite{cite_41}, the authors presented a noninvasive techniques to study the electronic properties of bilayer graphene using the angle resolved photoemission spectroscopy (ARPES) and the bilayer films was initially synthesized on a SiC substrate at the sufficiently low temperatures and at the given doping level. As the dopant's electronic states are situated far from the Fermi level or Dirac crossing energy in graphene films and due to the frozen states of the dopant's atoms the BLG structure could be considered as really suspended and independent from the substrate.  
      
This work presents a detailed analysis of the electronic and excitonic properties of suspended AA-stacked bilayer graphene when influenced by in-plane top and back layer voltages and doping. The methodology involves independent application of varying electric field potentials to the top and bottom layers leading to the resulting finite electric field between the layers. Experimentally, such structure of locally applied layer voltages was considered in Ref.\cite{cite_42} where the fabrication of local contacts has been performed using electron beam lithography. Another possibility to achieve experimentally the dual electric field potentials applied to each layer is to use the techniques of dual-gate device discussed in Refs. \cite{cite_43, cite_44, cite_45}. Such a techniques will consist of a planar back gate and o point like top gate (with the help of scanning tunneling microscope tip) applied to each single layers if the BLG structure. Our key findings include the coexistence of the antiferromagnetic and excitonic phases and the establishment of a significant bandgap in the electronic spectrum by tuning layer voltages, applicable to both doped and undoped bilayer graphene. For doped bilayer, our study reveals that applying substantial magnetic fields can lead to Wigner localized-like states, maintaining a charge imbalance between the sublattice layers that can be adjusted with layer voltage changes.

Moreover, for different values of applied voltages, we show the possibility for various states such as commensurate antiferromagnetic order, excitonic insulator, semiconducting and insulating.
 
We emphasizes the role of electronic correlation effects that facilitate the formation of excitons, which can be tailored by altering layer voltages and the interlayer Coulomb interaction leading to the spin-controllable formation of such quasiparticles. 

The paper is structured into Section \ref{sec:Section_2} that include the introduction of Hamiltonian governing the electronic correlations and external fields, Section \ref{sec:Section_3} with comprehensive analysis of numerical results, Section \ref{sec:Section_4} with summary of significant findings, and an Appendix \ref{sec:Section_5} discussing the calculation of thermodynamic averages. 
%
\section{\label{sec:Section_2} The model}
%
\subsection{\label{sec:Section_2_1} Hamiltonian of the system}
\begin{figure}[h!]
	\begin{center}
		\includegraphics[scale=0.6]{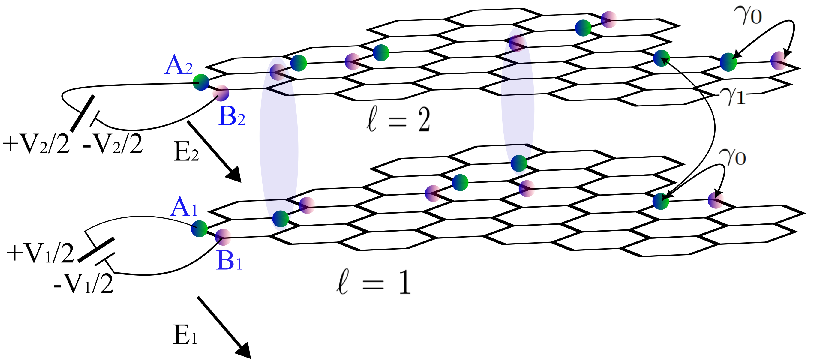}
		\caption{\label{fig:Fig_1} The presented structure of the gated AA-stacked bilayer graphene (AA-BLG) system illustrates the application of electric field potentials $V_1$ and $V_2$ to the bottom layer ($\ell = 1$ ) and top layer ($\ell = 2$), respectively. The diagram highlights the positions of atoms corresponding to the different sublattices: $A_1$ and $B_1$ in the bottom layer, and $A_2$ and $B_2$ in the top layer. In the diagram, the intralayer hopping parameter $\gamma_0$ represents the hopping between atoms within the same layer, while the interlayer hopping parameter $\gamma_1$ describes the tunneling between atoms of adjacent layers. Additionally, the formation of excitons is depicted in brown, illustrating the coupling between particles on sublattice sites $A_1$ and $A_2$, as well as $B_1$ and $B_2$. This excitonic formation occurs due to the electronic correlations.}
	\end{center}
\end{figure}

%
We consider the AA stacked bilayer graphene system. This setup allows for independent tuning of the potential on each sublattice, providing a versatile mechanism to control the electronic properties of the bilayer graphene system. The resulting electrostatic configuration significantly influences the band structure and excitonic characteristics, as illustrated in the schematic representation shown in Fig.~\ref{fig:Fig_1}. By adjusting the applied layer voltages $V_1$ and $V_2$, we can explore the effects of electronic correlations and interactions, leading to insights into the fundamental properties of the material under varying external conditions.
%
%
Additionally, a magnetic field is applied to the AA BLG system, perpendicular to the layers.
The model Hamiltonian of the system contains a tight-binding part and an interaction part. The tight-binding part is
\begin{eqnarray}
{\cal{\hat{H}}}_{\rm 0}&&=-\gamma_0\sum_{\left\langle {\bf{r}}{\bf{r}}'\right\rangle, \sigma}\left(\hat{a}^{\dag}_{1\sigma}\left({\bf{r}}\right)\hat{b}_{1\sigma}\left({\bf{r}}'\right)+\hat{a}^{\dag}_{2\sigma}\left({\bf{r}}\right)\hat{b}_{2\sigma}\left({\bf{r}}'\right)+h.c.\right)
	\nonumber\\
&&-\gamma_1\sum_{{\bf{r}},\sigma}\left(\hat{a}^{\dag}_{1\sigma}\left({\bf{r}}\right)\hat{a}_{2\sigma}\left({\bf{r}}\right)+\hat{b}^{\dag}_{1\sigma}\left({\bf{r}}\right)\hat{b}_{2\sigma}\left({\bf{r}}\right)+h.c.\right)
	\nonumber\\
	&&-\mu\sum_{\substack{{\bf{r}},\eta=a_{1},b_{1}\\ a_{2},b_{2}}}\hat{n}_{\eta}\left({\bf{r}}\right),
	\label{Equation_1}
\end{eqnarray}
and the interaction part is
\begin{eqnarray}
&&{\cal{\hat{H}}}_{\rm int}=U\sum_{\substack{\eta=a_{1},b_{1}\\ a_{2},b_{2}}}\sum_{{\bf{r}}}\hat{n}_{\eta\uparrow}\left({\bf{r}}\right)\hat{n}_{\eta\downarrow}\left({\bf{r}}\right)+
\nonumber\\
&&+W\sum_{{\bf{r}}\sigma\sigma'}\hat{n}_{1\sigma}\left({\bf{r}}\right)\hat{n}_{2\sigma'}\left({\bf{r}}\right)
	\nonumber\\
	&&+\frac{V_{1}}{2}\sum_{{\bf{r}}}\left(\hat{n}_{a1}\left({\bf{r}}\right)-\hat{n}_{b1}\left({\bf{r}}\right)\right)
	\nonumber\\
	&&+\frac{V_{2}}{2}\sum_{{\bf{r}}}\left(\hat{n}_{a2}\left({\bf{r}}\right)-\hat{n}_{b2}\left({\bf{r}}\right)\right)
	\nonumber\\
	&&-g\mu_{\rm B}B\sum_{{\bf{r}}\sigma}\sum_{\substack{\eta=a_{1},b_{1}\\ a_{2},b_{2}}}\xi_{\sigma}\hat{n}_{\eta\sigma}\left({\bf{r}}\right).
	\label{Equation_2}
\end{eqnarray}
In the aforementioned model Hamiltonian, the electron creation and annihilation operators $a^{\dag}_{\ell\sigma}\left({\bf{r}}\right)$, $b^{\dag}_{\ell\sigma}\left({\bf{r}}\right)$ and $a_{\ell\sigma}\left({\bf{r}}\right)$, $b_{\ell\sigma}\left({\bf{r}}\right)$ with $\ell=1,2$ are employed to describe the electronic states at the atomic lattice sites corresponding to the atomic positions $A_1, B_1$ in layer $1$, and $A_2, B_2$ in layer $2$, as depicted in Fig.~\ref{fig:Fig_1}.
The operators $\hat{n}_{\eta}\left({\bf{r}}\right)$ are defined as the spin-summed electron density operators for the sublattice electrons $\eta$, which can be represented as:
The operators $\hat{n}_{\eta}\left({\bf{r}}\right)$ are defined as the spin-summed electron density operators for the sublattice electrons $\eta$, which can be represented as:
\begin{eqnarray}
\hat{n}_{\eta}\left({\bf{r}}\right)=\sum_{\sigma}\hat{\eta}^{\dag}_{\sigma}\left({\bf{r}}\right)\hat{\eta}_{\sigma}\left({\bf{r}}\right),
\label{Equation_3}
\end{eqnarray}
where $\eta$ refers to the various sublattices, specifically: $\eta = a_{1}$ for site $A_1$ in layer $1$, $\eta = b_{1}$ for site $B_1$ in layer $1$, $\eta = a_{2}$ for site $A_2$ in layer $2$, $\eta = b_{2}$ for site $B_2$ in layer $2$.
The density operators in Eq.(\ref{Equation_3}) provide insight into the occupancy of the different sublattice sites.
%
\begin{figure}[h!]
	\begin{center}
		\includegraphics[scale=0.5]{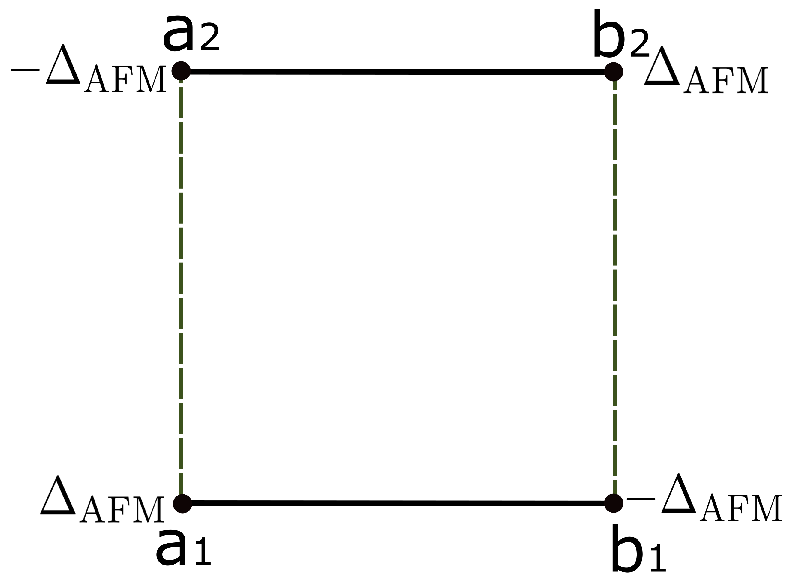}
		\caption{\label{fig:Fig_2} The staggering antiferromagnetic ordering in the AA bilayer graphene system. The antiferroamgnetic order parameter is shown at each sublattice site position in the bilayer. The order parameter $\Delta_{\rm AFM}$ is defined in Eq.(\ref{Equation_12}).}
	\end{center}
\end{figure} 
%
The first term in Eq.(\ref{Equation_1}) describes the intralayer hopping between the adjacent lattice sites within the same layer, characterized by the hopping parameter $\gamma_0$ and $\gamma_0\approx 2.8$ eV \cite{cite_46} in graphene layers. In our calculations, we will use the parameter $\gamma_0$ as the unit of energy. This choice simplifies our numerical evaluations and allows for clearer interpretation of the results, as all energies will be normalized relative to the characteristic hopping energy $\gamma_0$ in the AA-stacked bilayer graphene system.

The second term accounts for the interlayer hopping between corresponding sites of adjacent layers, described by the parameter $\gamma_1$.
The third term represents the chemical potential $\mu$, contributing to the energy of the system based on the electron density.

The first term in the interaction Hamiltonian in Eq.(\ref{Equation_2}) models the intra-sublattice Coulomb interaction with coupling strength $U$, accounting for electron-electron interactions on the same sublattice. The second term incorporates the inter-sublattice Coulomb interactions with coupling strength $W$. The next two terms $\frac{V_{1}}{2}$ and $\frac{V_{2}}{2}$ contribute to the potential energy from the applied electric fields acting on the sublattice densities $\hat{n}_{a1}$ and $\hat{n}_{b1}$ in layer $1$ and $\hat{n}_{a2}$ and $\hat{n}_{b2}$ in layer $2$.
The final term considers the interaction of the spin with the external magnetic field $B$, characterized by the parameters $g$ (Land\'e g-factor) (we take $g=2$ \cite{cite_47}) and $\mu_{\rm B}$ (Bohr magneton), along with the spin terms $\xi_{\sigma}$ which is defined as $\xi_{\sigma}=1$, if $\sigma=\uparrow$ and $\xi_{\sigma}=-1$, if $\sigma=\downarrow$. We, furthermore, investigate the consequences of these interactions on the electronic properties of the AA-stacked bilayer graphene, including the emergence of bandgaps excitonic pairing, the antiferromagnetic order parameter and the average charge distribution in the system. 

Indeed the binding energy of excitons in two-dimensional materials has been shown to be very strong, on the order of $1$ eV, owing to the significantly reduced charge carrier screening effects in atomically thin materials \cite{cite_48, cite_49}. Therefore, the optical properties of such low-dimensional systems should be dominated by excitons. For this reason, we have omitted screening effects whn considering interactions in our model. 
  
In the paper, we set $\hbar=1$, and $\mu_{\rm B}=1$. We will use the dimensionless magnetic field parameter $\tilde{B}=\mu_{\rm B}B/\gamma_0$. The intralayer hopping parameter $\gamma_0$ serves as the unit of energy scales in the problem. 

The operators $\hat{n}_{\eta\sigma}\left({\bf{r}}\right)$ in the interaction term are defined for each sublattice electrons as
\begin{eqnarray}
\hat{n}_{\eta\sigma}\left({\bf{r}}\right)=\hat{\eta}^{\dag}_{\sigma}\left({\bf{r}}\right)\hat{\eta}_{\sigma}\left({\bf{r}}\right).
\label{Equation_4}
\end{eqnarray}
The interaction terms $U$ and $W$ in Eq.(\ref{Equation_2}) can be rewritten in a more convenient form to facilitate the further linearization of the Hamiltonian. Specifically, for the $U$ term, we have:
\begin{eqnarray}
U\sum_{{\bf{r}},\eta}\hat{n}_{\eta\uparrow}\left({\bf{r}}\right)\hat{n}_{\eta\downarrow}\left({\bf{r}}\right)=\frac{U}{4}\left(\hat{n}^{2}_{\eta}\left({\bf{r}}\right)-\hat{p}^{2}_{z\eta}\left({\bf{r}}\right)\right),
\label{Equation_5}
\end{eqnarray}
where the operator $\hat{p}_{z\eta}\left({\bf{r}}\right)$ is the electron density polarization operator defined as:
\begin{eqnarray}
\hat{p}_{z\eta}\left({\bf{r}}\right)=\hat{n}_{\eta\uparrow}\left({\bf{r}}\right)-\hat{n}_{\eta\downarrow}\left({\bf{r}}\right).
\label{Equation_6}
\end{eqnarray}
Indeed, the quantity $\hat{p}_{z\eta}\left({\bf{r}}\right)$ is the polarization density corresponding to the differences in electron occupancy between the spin-up and spin-down states at site ${\bf{r}}$.
Next, transitioning to Grassmann variables for the electrons, as discussed in \cite{cite_50}, the $W$ interaction term in Eq.(\ref{Equation_2}) can be expressed as:
\begin{eqnarray}
&&W\sum_{{\bf{r}}\sigma\sigma'}{n}_{1\sigma}\left({\bf{r}}\right){n}_{2\sigma'}\left({\bf{r}}\right)=-W\sum_{{\bf{r}},\sigma\sigma'}|{\zeta}^{\left(a\right)}_{\sigma'\sigma}\left({\bf{r}}\right)|^{2}
\nonumber\\
&&-W\sum_{{\bf{r}},\sigma\sigma'}|{\zeta}^{\left(b\right)}_{\sigma'\sigma}\left({\bf{r}}\right)|^{2},
\label{Equation_7}
\end{eqnarray}
where the excitonic variables ${\zeta}^{(a)}_{\sigma'\sigma}\left({\bf{r}}\right)$, ${\zeta}^{(b)}_{\sigma'\sigma}\left({\bf{r}}\right)$, and their complex cojugates $\bar{\zeta}^{(a)}_{\sigma'\sigma}\left({\bf{r}}\right)$ and $\bar{\zeta}^{(b)}_{\sigma'\sigma}\left({\bf{r}}\right)$ are defined as:
\begin{eqnarray}
&&{\zeta}^{\left(a\right)}_{\sigma'\sigma}\left({\bf{r}}\right)=\bar{a}_{2\sigma'}\left({\bf{r}}\right){a}_{1\sigma}\left({\bf{r}}\right),
\nonumber\\
&&{\zeta}^{\left(b\right)}_{\sigma'\sigma}\left({\bf{r}}\right)=\bar{b}_{2\sigma'}\left({\bf{r}}\right){b}_{1\sigma}\left({\bf{r}}\right),
\nonumber\\
&&\bar{{\zeta}}^{\left(a\right)}_{\sigma'\sigma}\left({\bf{r}}\right)=\bar{a}_{1\sigma}\left({\bf{r}}\right){a}_{2\sigma'}\left({\bf{r}}\right),
\nonumber\\
&&\bar{{\zeta}}^{\left(b\right)}_{\sigma'\sigma}\left({\bf{r}}\right)=\bar{b}_{1\sigma}\left({\bf{r}}\right){b}_{2\sigma'}\left({\bf{r}}\right).
\label{Equation_8}
\end{eqnarray}
The total fermionic action of the system is defined as:
\begin{eqnarray}
{\cal{S}}={\cal{S}}_{\rm B}+\int^{\beta}_{0}d\tau\left({\cal{{H}}}_{\rm 0}\left(\tau\right)+{\cal{H}}_{\rm int}\left(\tau\right)\right),
\label{Equation_9}
\end{eqnarray}
where ${\cal{{H}}}_{\rm 0}$ and ${\cal{H}}_{\rm int}$ are Grassmann versions of the operators in Eqs.(\ref{Equation_1}) and (\ref{Equation_2}), and ${\cal{S}}_{\rm B}$, in Eq.(\ref{Equation_9}) is the fermionic Berry term defined for each sublattice fermions:
\begin{eqnarray}
{\cal{S}}_{\rm B}=\int^{\beta}_{0}d\tau\sum_{\substack{{\bf{r}},\eta=a_{1},b_{1}\\a_{2},b_{2}}}\bar{\eta}\left({\bf{r}}\tau\right)\partial_{\tau}\eta\left({\bf{r}}\tau\right),
\label{Equation_10}
\end{eqnarray}
and the partition function of the system will be
\begin{eqnarray}
{\cal{Z}}=\int\prod_{\substack{\eta=a_{1},b_{1}\\a_{2},b_{2}}}\left[{\cal{D}}\bar{\eta}{\cal{D}}{\eta}\right]e^{-{\cal{S}}}.
\label{Equation_11}
\end{eqnarray}
Next, we apply the Hubbard-Stratonovich decoupling procedure to linearize the biquadratic fermionic density terms in Eqs.(\ref{Equation_5}) and (\ref{Equation_7}), as detailed in our previous work \cite{cite_37}. The contributions to the fermionic action from the decoupling of the $U$ term in Eq.(\ref{Equation_5}), are
\begin{eqnarray}
{\cal{S}}_{\rm U}=\frac{U}{2}\int^{\beta}_{0}d\tau \sum_{{\bf{r}}\eta}n_{\eta}\left({\bf{r}}\tau\right)\left\langle n_{\eta}\left({\bf{r}}\tau\right)\right\rangle,
\nonumber\\
{\cal{S}}_{\Delta_{\rm p}}=-\frac{U}{2}\int^{\beta}_{0}d\tau \sum_{{\bf{r}}\eta}p_{z\eta}\left({\bf{r}}\tau\right)\left\langle p_{z\eta}\left({\bf{r}}\tau\right)\right\rangle.
\label{Equation_12}
\end{eqnarray}
Similarly, we can formulate the contribution arising from the decoupling of the interlayer interaction term in Eq.(\ref{Equation_7}) as follows:
\begin{eqnarray}
{\cal{S}}_{\Delta}=-\int^{\beta}_{0}d\tau \sum_{{\bf{r}}\sigma}\left(\Delta^{\left(a\right)}_{\sigma\sigma}\bar{\zeta}^{\left(a\right)}_{\sigma\sigma}\left({\bf{r}}\tau\right)+\right.
\nonumber\\
\left.+\Delta^{\left(b\right)}_{\sigma\sigma}\bar{\zeta}^{\left(b\right)}_{\sigma\sigma}\left({\bf{r}}\tau\right)
+c.c.\right).
\label{Equation_13}
\end{eqnarray}
In this expression, the parameters $\Delta^{(a)}_{\sigma\sigma}$, $\Delta^{(b)}_{\sigma\sigma}$ and their complex conjugates $\bar{\Delta}^{\left(a\right)}_{\sigma\sigma}, \bar{\Delta}^{\left(b\right)}_{\sigma\sigma}$ represent the excitonic order parameters, defined as:
\begin{eqnarray}
\Delta^{\left(a\right)}_{\sigma\sigma}=W\left\langle\zeta^{\left(a\right)}_{\sigma\sigma}\left({\bf{r}}\tau\right)\right\rangle,
\nonumber\\
\Delta^{\left(b\right)}_{\sigma\sigma}=W\left\langle\zeta^{\left(b\right)}_{\sigma\sigma}\left({\bf{r}}\tau\right)\right\rangle,
\nonumber\\
\bar{\Delta}^{\left(a\right)}_{\sigma\sigma}=W\left\langle\bar{\zeta}^{\left(a\right)}_{\sigma\sigma}\left({\bf{r}}\tau\right)\right\rangle,
\nonumber\\
\bar{\Delta}^{\left(b\right)}_{\sigma\sigma}=W\left\langle\bar{\zeta}^{\left(b\right)}_{\sigma\sigma}\left({\bf{r}}\tau\right)\right\rangle.
\label{Equation_14}
\end{eqnarray}
The parenthesis $\left\langle ... \right\rangle$, in Eqs.(\ref{Equation_12}) and (\ref{Equation_14}) are thermodynamic averages defined with the help of the partition function in Eq.(\ref{Equation_11}). We have
\begin{eqnarray}
\left\langle ... \right\rangle=\frac{1}{{\cal{Z}}}\int\prod_{\substack{\eta=a_{1},b_{1}\\a_{2},b_{2}}}\left[{\cal{D}}\bar{\eta}{\cal{D}}{\eta}\right] ... e^{-{\cal{S}}'},
\label{Equation_15}
\end{eqnarray}
where ${\cal{S}}'$ in Eq.(\ref{Equation_15}) is the fermionic action after all decoupling contributions (obtained in Eqs.(\ref{Equation_12}) and (\ref{Equation_13})):
\begin{eqnarray}
{\cal{S}}'=&&\int^{\beta}_{0}d\tau{\cal{H}}_{\rm 0}\left(\tau\right)+{\cal{S}}_{\rm U}+{\cal{S}}_{\Delta_{\rm p}}+
\nonumber\\
&&+{\cal{S}}_{\Delta}+{\cal{S}}_{\rm V}+{\cal{S}}_{\rm m}+{\cal{S}}_{\rm B},
\label{Equation_16}
\end{eqnarray}
where
\begin{eqnarray}
{\cal{S}}_{\rm V}=\int^{\beta}_{0}d\tau \sum_{{\bf{r}}}\left[\frac{V_{1}}{2}\left({n}_{a1}\left({\bf{r}}\tau\right)-{n}_{b1}\left({\bf{r}},\tau\right)\right)+\right.
\nonumber\\
\left.+\frac{V_{2}}{2}\left({n}_{a2}\left({\bf{r}}\tau\right)-{n}_{b2}\left({\bf{r}}\tau\right)\right)\right],
\label{Equation_17}
\end{eqnarray}
and 
\begin{eqnarray}
{\cal{S}}_{\rm m}=-g\mu_{\rm B}B\int^{\beta}_{0}d\tau \sum_{{\bf{r}}\sigma}\sum_{\substack{\eta=a_{1},b_{1}\\a_{2},b_{2}}}\xi_{\sigma}\hat{n}_{\eta\sigma}\left({\bf{r}}\right).
\label{Equation_18}
\end{eqnarray}
For simplification, we assume a homogeneous distribution of the gap parameter across the system and that these parameters are real. Therefore, we can express:
\begin{eqnarray}
\Delta^{\left(a\right)}_{\sigma\sigma}=\Delta^{\left(b\right)}_{\sigma\sigma}=\bar{\Delta}^{\left(a\right)}_{\sigma\sigma}=\bar{\Delta}^{\left(b\right)}_{\sigma\sigma}\equiv \Delta_{\sigma}.
\label{Equation_19}
\end{eqnarray} 
The index $\sigma$ represents the spins of the particles involved in the pairing, as is customary. In our case, $\sigma$ can take the values $\uparrow$ or $\downarrow$.
Additionally, we introduce the antiferromagnetic (AFM) order parameter for each sublattice, defined as follows based on the second contribution term in Eq. (\ref{Equation_12}):
\begin{eqnarray}
\Delta^{\eta}_{\rm AFM}=\frac{U}{2}\left\langle p_{z\eta}\left({\bf{r}}\tau\right)\right\rangle.
\label{Equation_20}
\end{eqnarray}
In a staggered configuration, we expect the antiferromagnetic order parameter to alternate in sign across the layers and sublattices, which can be represented as:
\begin{eqnarray}
\Delta^{a_{1}}_{\rm AFM}=-\Delta^{b_{1}}_{\rm AFM}=\Delta^{b_{2}}_{\rm AFM}=-\Delta^{a_{2}}_{\rm AFM}\equiv\Delta_{\rm AFM}.
\label{Equation_21}
\end{eqnarray}
This staggered configuration is illustrated in Fig. \ref{fig:Fig_2} and sometimes is called as the $G$-type antiferromagnetic ordering \cite{cite_51}. The staggered antiferromagnetic ordering in the AA bilayer graphene system is characterized by the organization of the antiferromagnetic order parameter at each sublattice site position within the bilayer structure. This ordering arises due to the electron correlations and the specific interlayer interactions that are present. The ordering pattern in Eq.(\ref{Equation_21}) enhances the stability of the system and contributes to the unique electronic and magnetic characteristics of the AA-stacked bilayer graphene, making it a fascinating subject of study in condensed matter physics.

In Section \ref{sec:Section_2_2}, we derive analytical expressions for the energies of the electronic band structure and establish a system of self-consistent equations to evaluate the important physical quantities, including: chemical potential, average charge density imbalances between the sublattice sites, antiferromagnetic order parameter, and excitonic order parameters.
These findings contribute to our understanding of the interplay between electronic correlations, magnetic order, and excitonic phenomena in the AA-stacked bilayer graphene system, especially under the influence of external electric and magnetic fields.
%
\subsection{\label{sec:Section_2_2} The energy spectrum}
%
We introduce the average electron density differences between the sublattice atomic sites $\delta\bar{n}_{1}$ for layer $\ell = 1$ and $\delta\bar{n}_{2}$ for layer $\ell = 2$:
\begin{eqnarray}
\bar{n}_{a1}-\bar{n}_{b1}=\delta{\bar{n}}_{1},
\nonumber\\
\bar{n}_{a2}-\bar{n}_{b2}=\delta{\bar{n}}_{2}.
\label{Equation_22}
\end{eqnarray}
Here and below, the bar notation $\bar{n}_{\eta}$ means the same statistical average as in Eq.(\ref{Equation_15}), i.e., $\bar{n}_{\eta}=\left\langle n_{\eta}\right\rangle$.
For both layers, we assume the same average electron occupation number $\kappa$, indicating equal electron doping:
\begin{eqnarray}
\bar{n}_{a1}+\bar{n}_{b1}=\bar{n}_{a2}+\bar{n}_{b2}=\frac{1}{\kappa}.
\label{Equation_23}
\end{eqnarray}
The value $\kappa = 0.5$ corresponds to half-filling, which represents the situation where the maximum allowable number of electrons at each atomic site is 2. Hence, $\kappa_{\text{min}} = 0.5$. Values $\kappa < \kappa_{\text{min}}$ are excluded due to the violation of the Pauli exclusion principle. Conversely, when $\kappa > 0.5$, it indicates a doped bilayer case, and we define the doping in the system as:
\begin{eqnarray}
x=\frac{1}{\kappa_{\rm min}}-\frac{1}{\kappa}. 
\label{Equation_24}
\end{eqnarray}
For instance, if $\kappa \in \left[0.5,1.0\right]$, the electron doping $x$ would then lie in the range $x\in\left[0, 1.0\right]$.
Transitioning to a Fourier space representation, the total fermionic action of the system can be expressed using the inverse Green’s function matrix ${\cal{G}}^{-1}_{{\bf{k}}\sigma}\left(\nu_{n}\right)$, defined for each spin direction separately. For this, we introduced Gorkov spinors $\bar{\Psi}$ and $\Psi$, pertinent to our analysis:
\begin{eqnarray} 
		{\Psi}_{{\bf{k}}\sigma}(\nu_{n})=\left(
	\begin{array}{crrrr}
		{a}_{1{\bf{k}}\sigma}(\nu_{n})\\\\
		{b}_{1{\bf{k}}\sigma}(\nu_{n}) \\\\
		{a}_{2{\bf{k}}\sigma}(\nu_{n}) \\\\
		{b}_{2{\bf{k}}\sigma}(\nu_{n}) \\\\
	\end{array}
	\right).
	\label{Equation_25}
\end{eqnarray}
The corresponding complex conjugate spinor is denoted as $\bar{\Psi}_{\sigma}\left({\bf{k}}, \nu_n\right)=\left(\bar{a}_{1{\bf{k}}\sigma}(\nu_{n}),
\bar{b}_{1{\bf{k}}\sigma}\left(\nu_{n}\right), \bar{a}_{2{\bf{k}}\sigma}(\nu_{n}\right),\bar{b}_{2{\bf{k}}\sigma}\left(\nu_{n}\right)$. Then the action of the system could be written as:
\begin{eqnarray}
{\cal{S}}\left[\bar{\Psi},\Psi\right]=\frac{1}{\beta{N}}\sum_{{\bf{k}}\nu_{n},\sigma}\bar{\Psi}_{{\bf{k}}\sigma}\left(\nu_{n}\right){\cal{G}}^{-1}_{{\bf{k}}\sigma}\left(\nu_{n}\right)\Psi_{{\bf{k}}\sigma}\left(\nu_{n}\right),
\label{Equation_26}
\end{eqnarray}
where the summation index ${\bf{k}}$ denotes the wave vector in reciprocal space, and $\nu_n$ are the fermionic Matsubara frequencies given by $\nu_n = \pi {k_B}T(2n + 1)/\hbar$ with $n$ being integers. The parameter $N$ represents the number of lattice sites in the graphene layers. 
%
\subsection{\label{sec:Section_2_3} Inverse Green's function matrix}
%
For the spin direction $\sigma = \uparrow$, the inverse Green's function matrix ${\cal{G}}^{-1}_{{\bf{k}}\uparrow}\left(\nu_n\right)$ is defined as:
\begin{eqnarray}
\noindent 
	\resizebox{\linewidth}{!}{\columnsep=1pt %
	${\cal{G}}^{-1}_{{\bf{k}}\uparrow}\left(\nu_{n}\right)=\left(
     \begin{matrix}
		-i\nu_{n}-\mu_{0\uparrow}-\Delta'_{1} & -\tilde{\gamma}_{{\bf{k}}} & -\gamma_1-\Delta_{\uparrow} & 0\\
		-{\tilde{\gamma}_{{\bf{k}}}}^{\ast} &-i\nu_{n}-\mu_{0\uparrow}+\Delta'_{1} &  0 & -\gamma_1-\Delta_{\uparrow}\\
		-\gamma_1-\bar{\Delta}_{\uparrow} & 0 & -i\nu_{n}-\mu_{0\uparrow}+\Delta'_{2} & -{\tilde{\gamma}_{{\bf{k}}}} \\
		0 & -\gamma_1-\bar{\Delta}_{\uparrow} & -{\tilde{\gamma}_{{\bf{k}}}}^{\ast} & -i\nu_{n}-\mu_{0\uparrow}-\Delta'_{2}\\
	\end{matrix}\right)$},
	\nonumber\\
	\label{Equation_27}
\end{eqnarray}
where we have defined
\begin{eqnarray} 
&&\mu_{0\uparrow}=\mu-\frac{U}{4\kappa}+g\mu_{\rm B}B,
\nonumber\\
&&\Delta'_{1}=\Delta_{1}-\frac{U\delta{\bar{n}}_{1}}{4},
\nonumber\\
&&\Delta'_{2}=\Delta_{2}+\frac{U\delta{\bar{n}}_{2}}{4}.
\nonumber\\
&&\Delta_{1}=\Delta_{\rm AFM}-\frac{V_{1}}{2},
\nonumber\\
&&\Delta_{2}=\Delta_{\rm AFM}+\frac{V_2}{2}.
\nonumber\\
\label{Equation_28}
\end{eqnarray}
The parameter $\tilde{\gamma}_{{\bf{k}}}$ is the dispersion in graphene layers
\begin{eqnarray} 
\tilde{\gamma}_{{\bf{k}}}=\gamma_0\left(e^{-ik_{x}a_{0}}+2e^{i\frac{k_{x}a_{0}}{2}}\cos\left(\frac{k_{y}a_{0}\sqrt{3}}{2}\right)\right).
\label{Equation_29}
\end{eqnarray}
We solve the equation $\det{{\cal{G}}^{-1}_{{\bf{k}}\uparrow}\left(\nu_{n}\right)}=0$ and we find the electronic band structure energies
\begin{eqnarray}
\varepsilon_{i{\bf{k}}\uparrow}=\mu_{0\uparrow}-\left(-1\right)^{i+1}\sqrt{A_{{\bf{k}}\uparrow}-\frac{1}{2}\sqrt{\zeta_{{\bf{k}}\uparrow}}},
\nonumber\\
\varepsilon_{j{\bf{k}}\uparrow}=\mu_{0\uparrow}-\left(-1\right)^{j+1}\sqrt{A_{{\bf{k}}\uparrow}+\frac{1}{2}\sqrt{\zeta_{{\bf{k}}\uparrow}}},
\label{Equation_30}
\end{eqnarray}
where $i=1,2$ and $j=3,4$. The quasienergy quantities $A_{{\bf{k}}\uparrow}$ and $\zeta_{{\bf{k}}\uparrow}$ are defined as
\begin{eqnarray}
&&A_{{\bf{k}}\uparrow}=|\tilde{\gamma}_{{\bf{k}}}|^{2}+|\Delta_{\uparrow}+\gamma_1|^{2}+\frac{\Delta'^{2}_{1}+\Delta'^{2}_{2}}{2}
\nonumber\\
&&\zeta_{{\bf{k}}\uparrow}=\left(\Delta'^{2}_{1}-\Delta'^{2}_{2}\right)^{2}+4|\Delta_{\uparrow}+\gamma_1|^{2}\times
\nonumber\\
&&\times\left[\left(\Delta'_{1}-\Delta'_{2}\right)^{2}+4|\tilde{\gamma}_{{\bf{k}}}|^{2}\right].
\nonumber\\
\label{Equation_31}
\end{eqnarray}
The inverse Green's function matrix for the spin direction $\sigma=\downarrow$ is
\begin{eqnarray}
\noindent 
	\resizebox{\linewidth}{!}{\columnsep=1pt %
	${\cal{G}}^{-1}_{{\bf{k}}\downarrow}\left(\nu_{n}\right)=\left(
     \begin{matrix}
		-i\nu_{n}-\mu_{0\downarrow}-\Delta'_{3} & -\tilde{\gamma}_{{\bf{k}}} & -\gamma_1-\Delta_{\downarrow} & 0\\
		-{\tilde{\gamma}_{{\bf{k}}}}^{\ast} &-i\nu_{n}-\mu_{0\downarrow}+\Delta'_{3} &  0 & -\gamma_1-\Delta_{\downarrow}\\
		-\gamma_1-\bar{\Delta}_{\downarrow} & 0 & -i\nu_{n}-\mu_{0\downarrow}+\Delta'_{4} & -{\tilde{\gamma}_{{\bf{k}}}} \\
		0 & -\gamma_1-\bar{\Delta}_{\downarrow} & -{\tilde{\gamma}_{{\bf{k}}}}^{\ast} & -i\nu_{n}-\mu_{0\downarrow}-\Delta'_{4}\\
	\end{matrix}\right)$},
	\nonumber\\
	\label{Equation_32}
\end{eqnarray}
where the parameters $\mu_{0\downarrow}$ and $\Delta'_{i}$ ($i=3,4$) for spin $\sigma = \downarrow$ are defined as follows:
\begin{eqnarray} 
&&\mu_{0\downarrow}=\mu-\frac{U}{4\kappa}-g\mu_{\rm B}B,
\nonumber\\
&&\Delta'_{3}=\Delta_{3}+\frac{U\delta{\bar{n}}_{1}}{4},
\nonumber\\
&&\Delta'_{4}=\Delta_{4}-\frac{U\delta{\bar{n}}_{2}}{4},
\nonumber\\
&&\Delta_{3}=\Delta_{\rm AFM}+\frac{V_{1}}{2},
\nonumber\\
&&\Delta_{4}=\Delta_{\rm AFM}-\frac{V_2}{2}.
\label{Equation_33}
\end{eqnarray}
To determine the electronic band structure, we solve the equation $\det{\cal{G}}^{-1}_{{\bf{k}}\downarrow}\left(\nu_n\right) = 0$ to obtain the energy eigenvalues corresponding to the spin-down states:
\begin{eqnarray}
\varepsilon_{i{\bf{k}}\downarrow}=\mu_{0\downarrow}-\left(-1\right)^{i+1}\sqrt{A_{{\bf{k}}\downarrow}-\frac{1}{2}\sqrt{\zeta_{{\bf{k}}\downarrow}}},
\nonumber\\
\varepsilon_{j{\bf{k}}\downarrow}=\mu_{0\downarrow}-\left(-1\right)^{j+1}\sqrt{A_{{\bf{k}}\downarrow}+\frac{1}{2}\sqrt{\zeta_{{\bf{k}}\downarrow}}},
\label{Equation_34}
\end{eqnarray}
where $i=1,2$ and $j=3,4$. The quasienergy quantities $A_{{\bf{k}}\downarrow}$ and $\zeta_{{\bf{k}}\downarrow}$ for the spin-down states are defined similarly to those for the spin-up states:
\begin{eqnarray}
&&A_{{\bf{k}}\downarrow}=|\tilde{\gamma}_{{\bf{k}}}|^{2}+|\Delta_{\downarrow}+\gamma_1|^{2}+\frac{\Delta'^{2}_{3}+\Delta'^{2}_{4}}{2}
\nonumber\\
&&\zeta_{{\bf{k}}\downarrow}=\left(\Delta'^{2}_{3}-\Delta'^{2}_{4}\right)^{2}+4|\Delta_{\downarrow}+\gamma_1|^{2}\times
\nonumber\\
&&\times\left[\left(\Delta'_{3}-\Delta'_{4}\right)^{2}+4|\tilde{\gamma}_{{\bf{k}}}|^{2}\right].
\nonumber\\
\label{Equation_35}
\end{eqnarray}
Next, we introduce external source terms and proceed to calculate the statistical averages from Eqs.(\ref{Equation_12}), (\ref{Equation_14}), (\ref{Equation_20}), (\ref{Equation_22}) and (\ref{Equation_23}) (for further details, please refere to Appendix \ref{sec:Section_5}). We obtain the following system of self-consistent equations that describe the behavior of the electron densities and order parameters within the AA-stacked bilayer graphene system:
\begin{widetext}
\begin{eqnarray}
&&\frac{1}{N}\sum_{{\bf{k}}}\sum^{4}_{i=1}\left(\alpha^{\uparrow}_{i{\bf{k}}}n_{\rm F}\left(\mu-\varepsilon_{i{\bf{k}}\uparrow}\right)+\alpha^{\downarrow}_{i{\bf{k}}}n_{\rm F}\left(\mu-\varepsilon_{i{\bf{k}}\downarrow}\right)\right)=\frac{2}{\kappa},
\nonumber\\
&&\frac{1}{N}\sum_{{\bf{k}}}\sum^{4}_{i=1}\left(\beta^{\uparrow}_{1i{\bf{k}}}n_{\rm F}\left(\mu-\varepsilon_{i{\bf{k}}\uparrow}\right)+\beta^{\downarrow}_{1i{\bf{k}}}n_{\rm F}\left(\mu-\varepsilon_{i{\bf{k}}\downarrow}\right)\right)=\delta{\bar{n}}_{1},
\nonumber\\
&&\frac{1}{N}\sum_{{\bf{k}}}\sum^{4}_{i=1}\left(\beta^{\uparrow}_{2i{\bf{k}}}n_{\rm F}\left(\mu-\varepsilon_{i{\bf{k}}\uparrow}\right)+\beta^{\downarrow}_{2i{\bf{k}}}n_{\rm F}\left(\mu-\varepsilon_{i{\bf{k}}\downarrow}\right)\right)=\delta{\bar{n}}_{2},
\nonumber\\
&&\frac{U}{N}\sum_{{\bf{k}}}\sum^{4}_{i=1}\left(\gamma^{\uparrow}_{i{\bf{k}}}n_{\rm F}\left(\mu-\varepsilon_{i{\bf{k}}\uparrow}\right)-\gamma^{\downarrow}_{i{\bf{k}}}n_{\rm F}\left(\mu-\varepsilon_{i{\bf{k}}\downarrow}\right)\right)=\Delta_{\rm AFM},
\nonumber\\
&&\frac{W(\Delta_{\uparrow}+\gamma_1)}{4N}\sum_{{\bf{k}}}\sum^{4}_{i=1}\delta^{\uparrow}_{i{\bf{k}}}n_{\rm F}\left(\mu-\varepsilon_{i{\bf{k}}\uparrow}\right)=\Delta_{\uparrow},
\nonumber\\
&&\frac{W(\Delta_{\downarrow}+\gamma_1)}{4N}\sum_{{\bf{k}}}\sum^{4}_{i=1}\delta^{\downarrow}_{i{\bf{k}}}n_{\rm F}\left(\mu-\varepsilon_{i{\bf{k}}\downarrow}\right)=\Delta_{\downarrow}.
\label{Equation_36}
\end{eqnarray}
\end{widetext}
In these equations, $n_{\rm F}\left(x\right)$ denotes the Fermi-Dirac distribution function, given by:
\begin{eqnarray}
n_{\rm F}\left(x\right)=\frac{1}{e^{\beta\left(x-\mu\right)}+1}.
\label{Equation_37}
\end{eqnarray}
The coefficients $\alpha^{\sigma}_{i{\bf{k}}}$, $\beta^{\sigma}_{ji{\bf{k}}}$, $\gamma^{\sigma}_{i{\bf{k}}}$, and $\delta^{\sigma}_{i{\bf{k}}}$ (where $j=1,2$ and $i=1,...,4$) are defined and calculated in Appendix \ref{sec:Section_5}. The system of equations in Eq. (\ref{Equation_36}) can be solved numerically. The complete solutions, considering varying interaction parameters, will be presented in the following section. This analysis will reveal the intricate interplay between the electronic properties, magnetic ordering, and excitonic behavior in the AA-stacked bilayer graphene under different conditions of external fields and interactions. 
%
\section{\label{sec:Section_3} Numerical solution}
%
\subsection{\label{sec:Section_3_1} A general discussion}
\begin{figure}[h!]
\begin{center}
\includegraphics[scale=0.32]{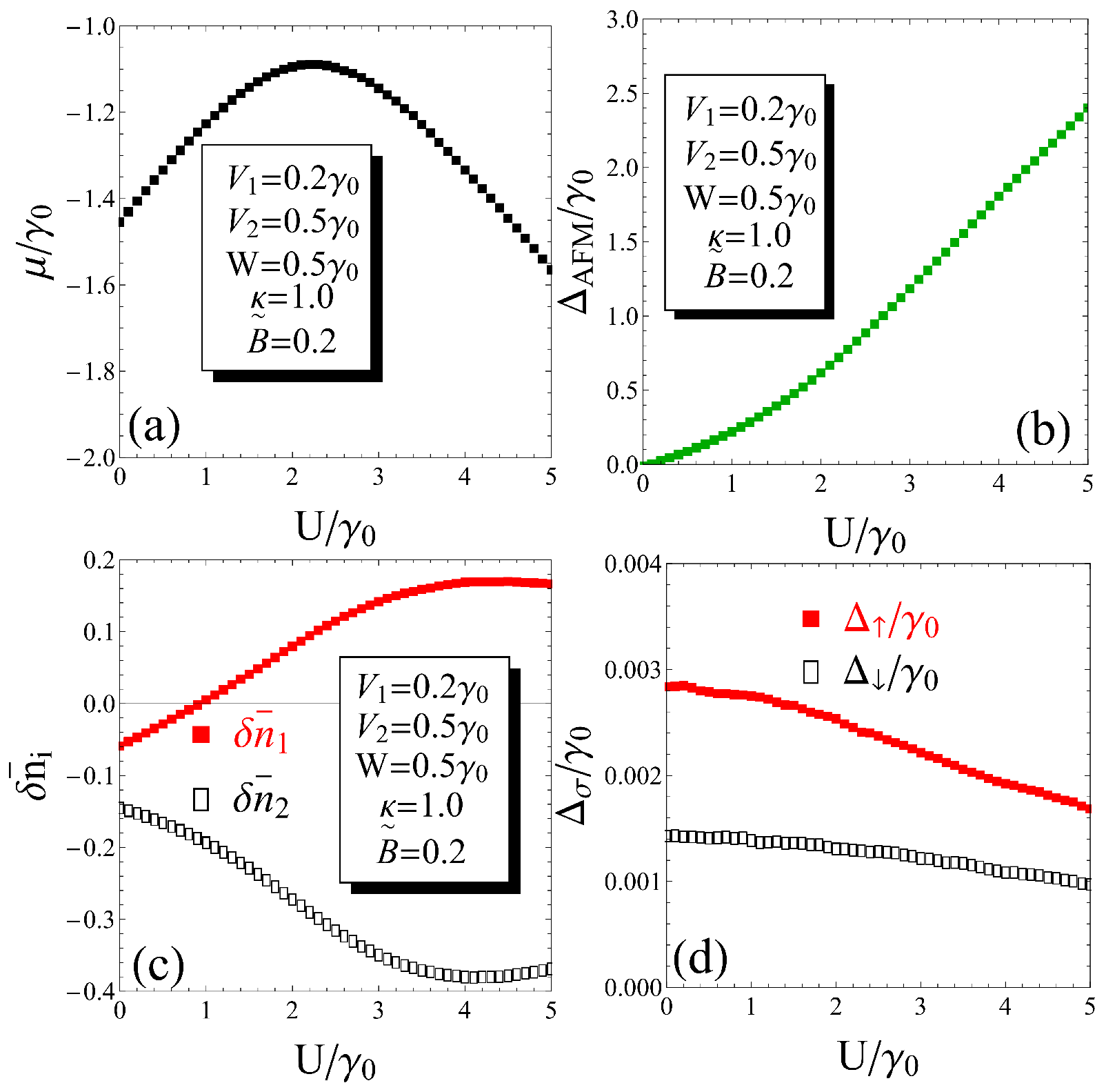}
\caption{\label{fig:Fig_3} The numerical solutions for key physical quantities are presented as functions of the local Coulomb potential $U$. Specifically, the following results are illustrated: panel (a): chemical potential $\mu$, panel (b): Antiferromagnetic order parameter $\Delta_{\rm AFM}$, panel (c): average density imbalance functions $\delta \bar{n}_{1}$ and $\delta \bar{n}_{2}$, panel (d): Excitonic order parameters $\Delta_{\uparrow}$ and $\Delta_{\downarrow}$. The interlayer Coulomb potential is fixed at $W = 0.5 \gamma_0$, while the magnetic field parameter $\tilde{B}$ is set to $\tilde{B} = 0.2$. Additionally, the layer voltages are set to $V_1 = 0.2 \gamma_0$ and $V_2 = 0.5 \gamma_0$.}
\end{center}
\end{figure}

%
In this section, we discuss the numerical results obtained following the procedure outlined in Eq. (\ref{Equation_36}) from Section \ref{sec:Section_2_3}. The calculations were conducted using the finite difference approximation, employing Newton's fast convergent algorithm \cite{cite_52} for efficient convergence. In Fig. \ref{fig:Fig_3}, we present the numerical results displaying the dependence of various physical quantities on the local Coulomb potential $U$: panel (a): Chemical potential $\mu$, panel (b): antiferromagnetic order parameter $\Delta_{\rm AFM}$, panel (c): average charge density difference functions $\delta \bar{n}_{1}$ and $\delta\bar{n}_{2}$, panel (d): excitonic order parameters $\Delta_{\uparrow}$ and $\Delta_{\downarrow}$.
In these calculations, the interlayer interaction potential is set to $W = 0.5 \gamma_0 = 1.4$ \text{eV}. The applied electric field potentials are fixed at $V_{1} = -0.5 \gamma_0 = -1.4$ \text{eV} and $V_{2}=0.5\gamma_0=1.4$ \text{eV}. The doped bilayer graphene has been considered in the calculations in Fig. \ref{fig:Fig_3} with $x=1.0$ (see in Eq.(\ref{Equation_24})). At the equilibrium state of the system, the local Coulomb potential $U$ plays a crucial role in stabilizing the charge density imbalance between sublattice sites. The strong electron repulsion induced by $U$ leads to localized electron occupancy at the lattice sites. This localization results in an enhancement of the charge density differences between the sublattices.

As the value of $U$ increases sufficiently, the charge density imbalances tend to stabilize and exhibit minimal variation, indicating that the system reaches a new equilibrium state where the effects of the local Coulomb interactions dominate.
As shown in panel (c), the functions $\delta \bar{n}_{i}$ stabilize at high values of the local Coulomb interaction, specifically around $U \approx 3.5 \gamma_0 = 9.8$ \text{eV}. This value coincides with that discussed in Ref. \cite{cite_53}, which relates to the formation of a gapped spin-liquid phase. Notably, $\delta \bar{n}_{i}$ remains nearly constant for $U \geq 4 \gamma_0$.

The antiferromagnetic order parameter is found to be $\Delta_{\rm AFM} = 1.508 \gamma_0 = 4.22$ \text{eV} at $U = 3.5 \gamma_0$ and $\Delta_{\rm AFM} \geq 1.82 \gamma_0$ for the range $U \geq 4 \gamma_0$. These obtained values align well with Hartree-Fock results presented in Refs. \cite{cite_54, cite_55}, as well as with Monte Carlo calculations in Ref. \cite{cite_56}.

Additionally, we observe that the excitonic order parameters $\Delta_{\uparrow}$ and $\Delta_{\downarrow}$ decrease as the local Coulomb potential $U$ increases. This behavior can be attributed to the localizing effect of the Hubbard $U$ potential, which enhances electron localization at the lattice sites. As a result, the formation of excitonic states is suppressed, leading to a reduction in the excitonic order parameters.

%
%
%
\begin{figure}[h!]
\includegraphics[scale=0.32]{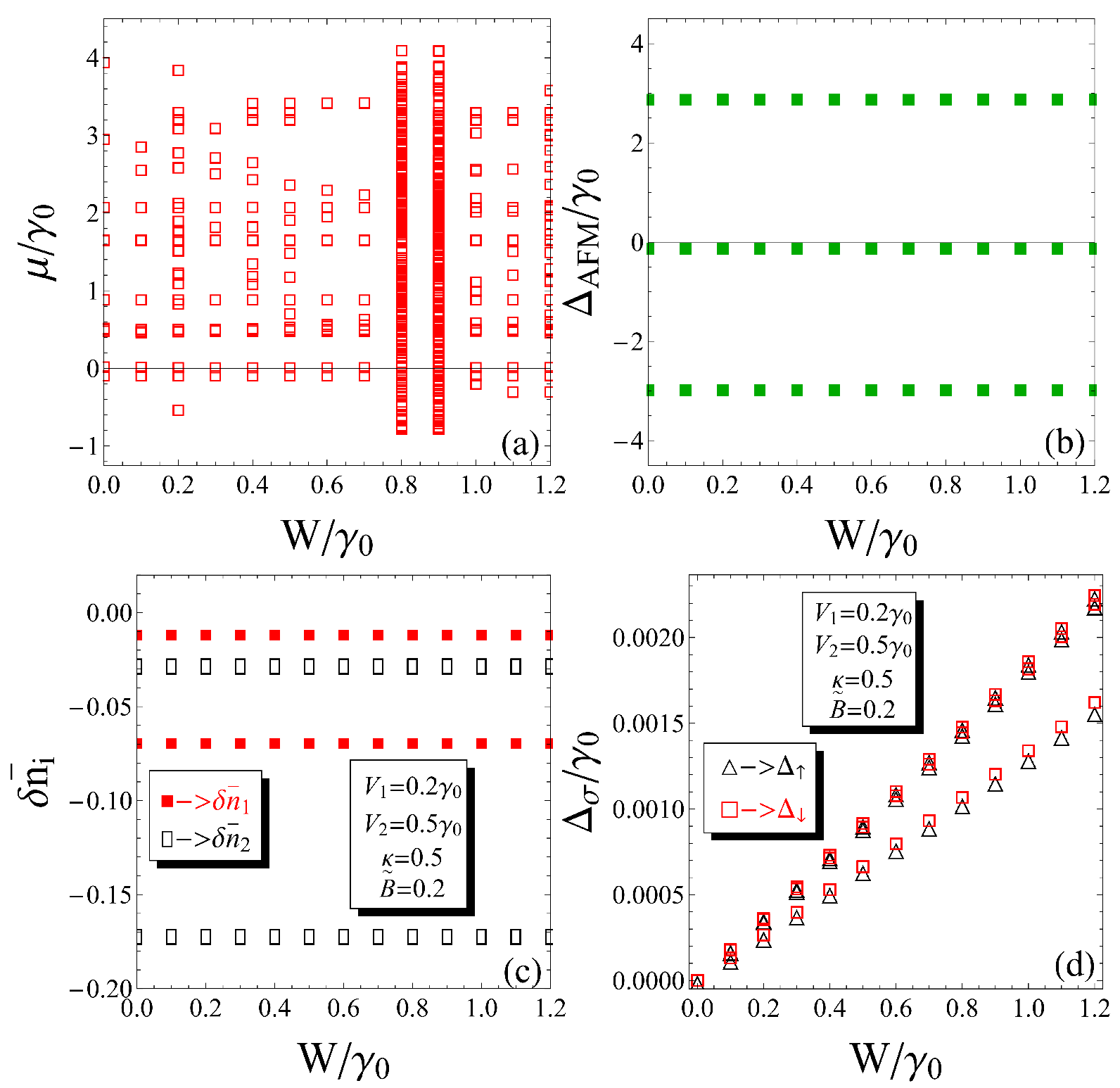}
\caption{\label{fig:Fig_4} The numerical solutions for key physical quantities are presented (in undoped AA bilayer graphene with $x=0.0$) as functions of the interlayer Coulomb potential $W$. Specifically, the following results are illustrated: panel (a): chemical potential $\mu$, panel (b): antiferromagnetic order parameter $\Delta_{\rm AFM}$, panel (c): average density imbalance functions $\delta \bar{n}_{1}$ and $\delta \bar{n}_{2}$, panel (d): excitonic order parameters $\Delta_{\uparrow}$ and $\Delta_{\downarrow}$. The intralayer Coulomb potential is fixed at $U = 3.3\gamma_0$, while the magnetic field parameter $\tilde{B}$ is set to $\tilde{B} = 0.2$. Additionally, the layer voltages are set to $V_1 = 0.2 \gamma_0=0.56$ \text{eV} and $V_2 = 0.5 \gamma_0=1.4$ \text{eV}.}
\end{figure} 
\begin{figure}[h!]
\includegraphics[scale=0.32]{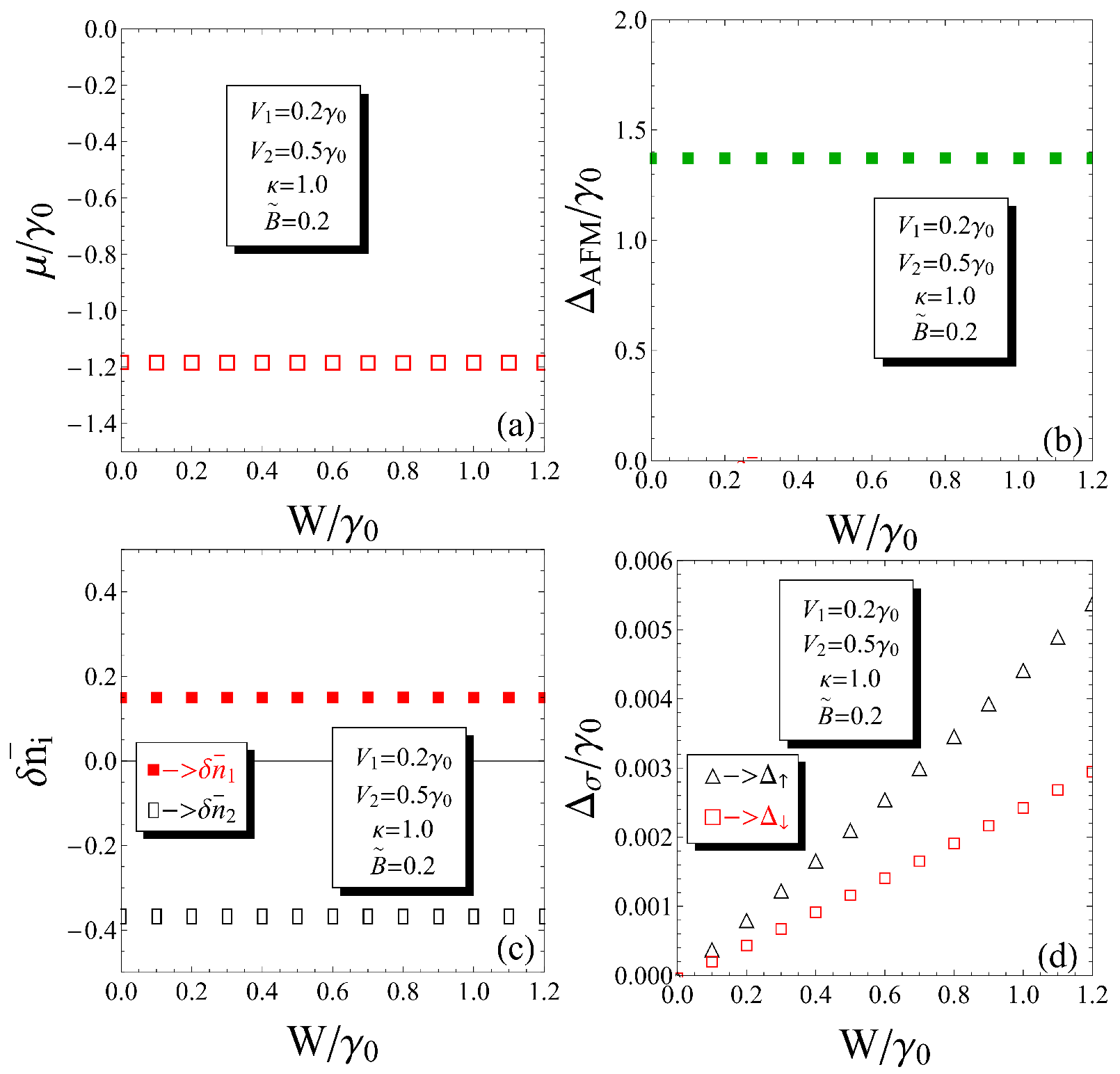}
\caption{\label{fig:Fig_5} The numerical solutions for key physical quantities (in doped AA bilayer graphene: $x=1$) are presented as functions of the interlayer Coulomb potential $W$. Specifically, the following results are illustrated: panel (a): chemical potential $\mu$, panel (b): antiferromagnetic order parameter $\Delta_{\rm AFM}$, panel (c): average density imbalance functions $\delta \bar{n}_{1}$ and $\delta \bar{n}_{2}$, panel (d): excitonic order parameters $\Delta_{\uparrow}$ and $\Delta_{\downarrow}$. The intralayer Coulomb potential is fixed at $U = 3.3\gamma_0$, while the magnetic field parameter $\tilde{B}$ is set to $\tilde{B} = 0.2$. Additionally, the layer voltages are set to $V_1 = 0.2 \gamma_0=0.56$ \text{eV} and $V_2 = 0.5 \gamma_0=1.4$ \text{eV}.}
\end{figure} 
%
In Figs.\ref{fig:Fig_4} and \ref{fig:Fig_5}, we present numerical calculations for the same physical quantities as in Fig.\ref{fig:Fig_3}, this time as a function of the interlayer Coulomb interaction parameter $W$, while maintaining a fixed value of $U = 3.3 \gamma_0$. All other parameters remain consistent with those in Fig.\ref{fig:Fig_3}. We examine two scenarios for the AA BLG:

\begin{enumerate}
\item \textbf{undoped case}: when $\kappa = 0.5$ (or $x = 0$). The results are shown in Fig. \ref{fig:Fig_4}.
\item \textbf{doped case}: for large doping with $\kappa = 1.0$ (or $x = 1.0$). The results are depicted in Fig. \ref{fig:Fig_5}.
\end{enumerate}

We observe significant differences in the solutions for these two cases. In the undoped scenario, the solutions of the chemical potential exhibit a stripe-like structure as a function of $W$ (see panel (a) in Fig. \ref{fig:Fig_4}). Conversely, for the doped case, shown in panel (a) of Fig. \ref{fig:Fig_5}, the chemical potential solutions appear single-valued for each $W$ value.

Additionally, the antiferromagnetic gap parameter $\Delta_{\rm AFM}$ presents a triple solution for each $W$ in the undoped case (see panel (b) in Fig. \ref{fig:Fig_4}), signifying that the undoped AA BLG system can exist in three distinct magnetic polarization states. In contrast, for the doped BLG case, the solutions for the gap parameter $\Delta_{\rm AFM}$ are single-valued, indicating that the system stabilizes in a single-polarized state.

Similar behaviors are observed in the average charge density differences $\delta \bar{n}_{i}$ and the excitonic order parameters $\Delta{\uparrow}$ and $\Delta_{\downarrow}$.

An interesting observation in the undoped AA BLG case is that transitioning between the upper magnetic polarized state $+\Delta_{\rm AFM}$ and the lower magnetic polarized state $-\Delta_{\rm AFM}$ results in the interchange of numerical values for the excitonic gap parameters with different spin orientations. Specifically, for the given antiferromagnetic polarization state, the excitonic formations are governed by two spin channels: for $+\Delta_{\rm AFM}$, we find the pair of solutions $\left(\Delta_{\uparrow}, \Delta_{\downarrow}\right)$, whereas for the state $-\Delta_{\rm AFM}$, the pairs switch to $\left(\Delta_{\downarrow}, \Delta_{\uparrow}\right)$.

In the case of doped AA BLG, we observe that the function $\delta\bar{n}_{1} > 0$ and $\delta\bar{n}_{2} < 0$ (shown in panel (c) of Fig. \ref{fig:Fig_5}). This indicates that there are more electrons on the $A_1$ sublattice sites compared to the $B_1$ lattice sites, and the reverse is true for the upper layer $\ell=2$. For the undoped case, shown in panel (c) of Fig. \ref{fig:Fig_4}, we find that $\delta \bar{n}_{i} < 0$ for both $i=1,2$. Moreover, a particularly surprising result in this case is that we get two possible negative values for each average charge occupation difference $\delta \bar{n}_{i}$, which contrasts with the doped BLG system, where we observe only a single numerical solution for each $\delta \bar{n}_{i}$ (see in Fig. \ref{fig:Fig_5}).

%
\subsection{\label{sec:Section_3_2} Magnetic field dependence}
%
\begin{figure}[h!]
\includegraphics[scale=0.31]{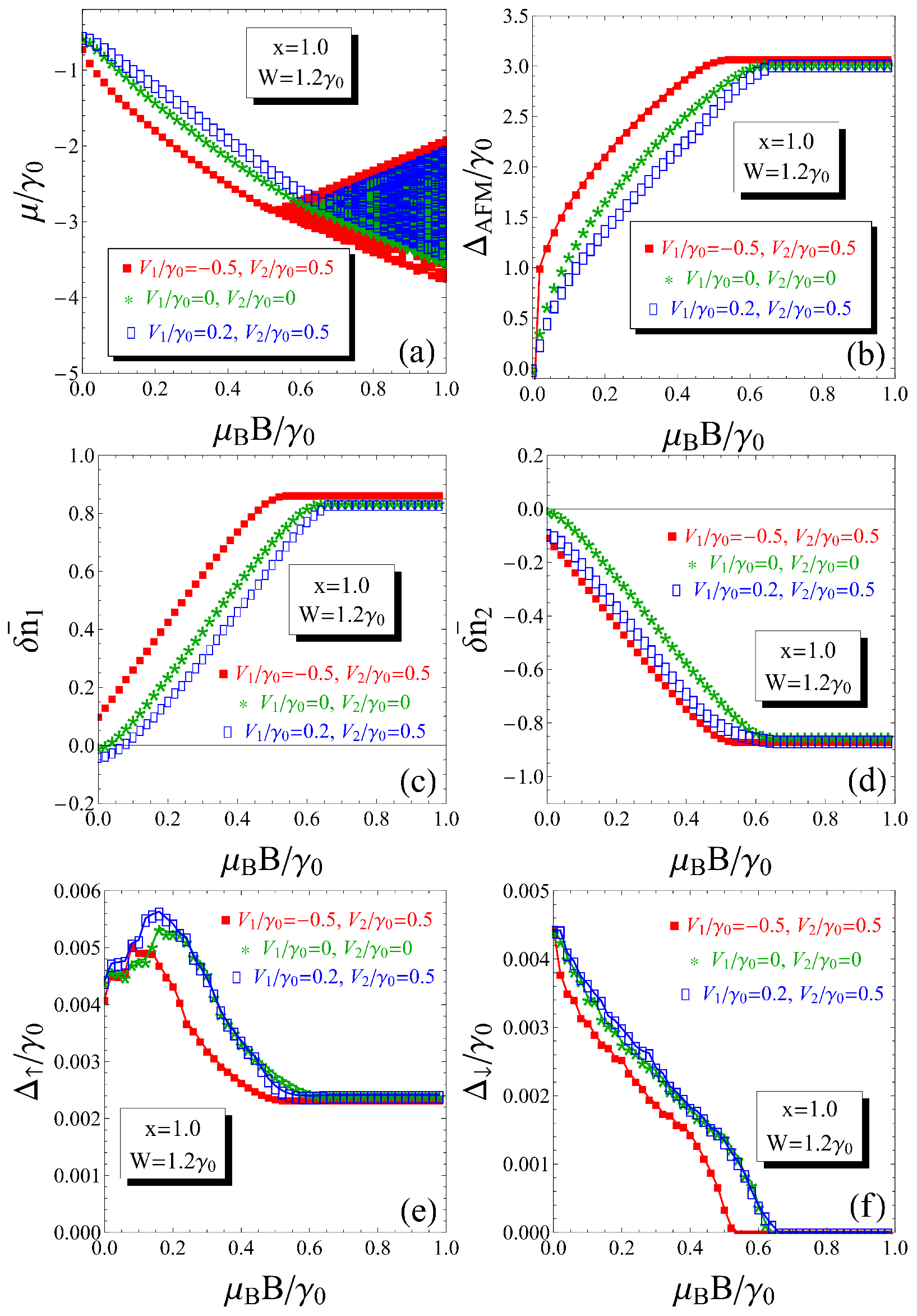}
\caption{\label{fig:Fig_6} The numerical solutions for key physical quantities (in doped AA bilayer graphene: $x=1$) are presented as functions of the external magnetic field parameter $\mu_{B}B/\gamma_0$. Specifically, the following results are illustrated: panel (a): chemical potential $\mu$, panel (b): antiferromagnetic order parameter $\Delta_{\rm AFM}$, panel (c): average density imbalance functions $\delta \bar{n}_{1}$ and $\delta \bar{n}_{2}$, panel (d): excitonic order parameters $\Delta_{\uparrow}$ and $\Delta_{\downarrow}$. The intralayer Coulomb potential is fixed at $U = 3.3\gamma_0$, while the interlayer Coulomb interaction parameter is $\tilde{B}$ fized at$W=1.2\gamma_0$. Different values of the layer voltages have been considered: $V_1 = -0.5 \gamma_0=-1.4$\text{eV} and $V_2 = 0.5 \gamma_0=1.4$ \text{eV}(see curves in Red), $V_1 = 0$ and $V_2 = 0$ (see curves in Green) and $V_1 = 0.2 \gamma_0=0.56$ \text{eV} and $V_2 = 0.5 \gamma_0=1.4$ \text{eV}(see curves in blue). The calculations have been done at large doping with $x=1.0$.}
\end{figure} 
%
In Fig.\ref{fig:Fig_6}, we present the magnetic field dependence of the calculated physical quantities. We consider different values of the electric field potentials, specifically setting $V_1 = -0.5 \gamma_0 = -1.4$ \text{eV} and $V_2 = 0.5 \gamma_0 = 1.4$ \text{eV} (represented by red curves), $V_1 = 0$ and $V_2 = 0$ (green curves), and $V_1 = 0.2 \gamma_0 = 0.56 \text{eV}$ and $V_2 = 0.5 \gamma_0 = 1.4$ \text{eV} (blue curves). The electron doping is set to a large value of $x = 1.0$.

A primary focus is on the behavior of the chemical potential calculated in panel (a) of Fig.\ref{fig:Fig_6}. Notably, we identify a critical value $\tilde{B}_{\rm C} = \mu_{B}B_C/\gamma_0$ of the magnetic field parameter $\mu_{B}B/\gamma_0$ at which the physical quantities exhibit a drastic change. For instance, in panel (a), the chemical potential shows a single solution for all values of $\tilde{B} < \tilde{B}_{\rm C}$, while it presents a band solution for $\tilde{B} > \tilde{B}_{\rm C}$. As illustrated, the positions of these critical values vary with changes in the external external electric field potentials $V_1$ and $V_2$.

The band solution observed in the chemical potential signifies that for each value of the external magnetic field, there are multiple possibilities for single-particle excitations within the system. Consequently, the system can select the appropriate value of $\mu$ to enable at least one single-particle excitation. The absolute value of $\mu$ represents the real energy required to excite a single particle from the system. When $|\mu|$ is small (as seen in the blue curve in panel (a)), more quasiparticles can be excited simultaneously, potentially forming excitons, as indicated in panels (e) and (f) of Fig. \ref{fig:Fig_6}.

In panels (e) and (f), in Fig.\ref{fig:Fig_6}, we observe that the magnitudes of the excitonic gap parameters $\Delta_{\uparrow}$ and $\Delta_{\downarrow}$ are larger when $|\mu|$ is small. The value of $|\mu|$ also serves as an indicator of localization within the system. This is evident in panel (b) of Fig. \ref{fig:Fig_6}, where we see that at small values of $|\mu|$, the antiferromagnetic order parameter is smaller (blue curve), while larger values of $|\mu|$ (red curves) correspond to a stronger antiferromagnetic order.

The splitting behavior of the chemical potential has also been observed for the undoped and vertically gated (single interlayer voltage) AA BLG structure, as discussed in Ref. \cite{cite_57}. In the doped case, however, only single solutions for $\mu$ were obtained. Thus, the splitting of $\mu$ can be attributed to the double-gated AA BLG system explored here. It is important to note that $\mu$ yields only single solutions (for both doped and undoped cases) in AB BLG systems, as demonstrated in Ref. \cite{cite_58}.

Furthermore, the antiferromagnetic order parameter increases with $\tilde{B}$ and reaches its maximum value at $\tilde{B} = \tilde{B}_{\rm C}$. This indicates that the magnetic field enhances the antiferromagnetic ground state until reaching a critical value. For $\tilde{B} > \tilde{B}_{\rm C}$, the antiferromagnetic order parameter becomes constant, denoted as $\Delta_{\rm AFM} \equiv \Delta_0$. The constant nature of the antiferromagnetic order influences the behavior of the average charge density imbalance functions $\delta \bar{n}_{1}$ and $\delta \bar{n}_{2}$, as calculated in panels (c) and (d) of Fig.\ref{fig:Fig_6}.

A surprising result emerges in the analysis of the excitonic gap parameters associated with different spin channels, $\Delta_{\uparrow}$ and $\Delta_{\downarrow}$. These results are depicted in panels (e) and (f) of Fig. \ref{fig:Fig_6}. Notably, the gap parameter $\Delta_{\uparrow}$ transitions to a constant value for $\tilde{B} > \tilde{B}_{\rm C}$, whereas the gap parameter $\Delta{\downarrow}$ vanishes in the same regime, for all considered values of the external electric field potentials $V_1$ and $V_2$.

This behavior indicates that the magnetic field plays a critical role in controlling the formation of excitonic states in different spin channels. Furthermore, it suggests that the magnetic field has the potential to modulate spin-dependent excitonic transport and energy transfers within the system. 

\begin{figure}[h!]
\includegraphics[scale=0.2]{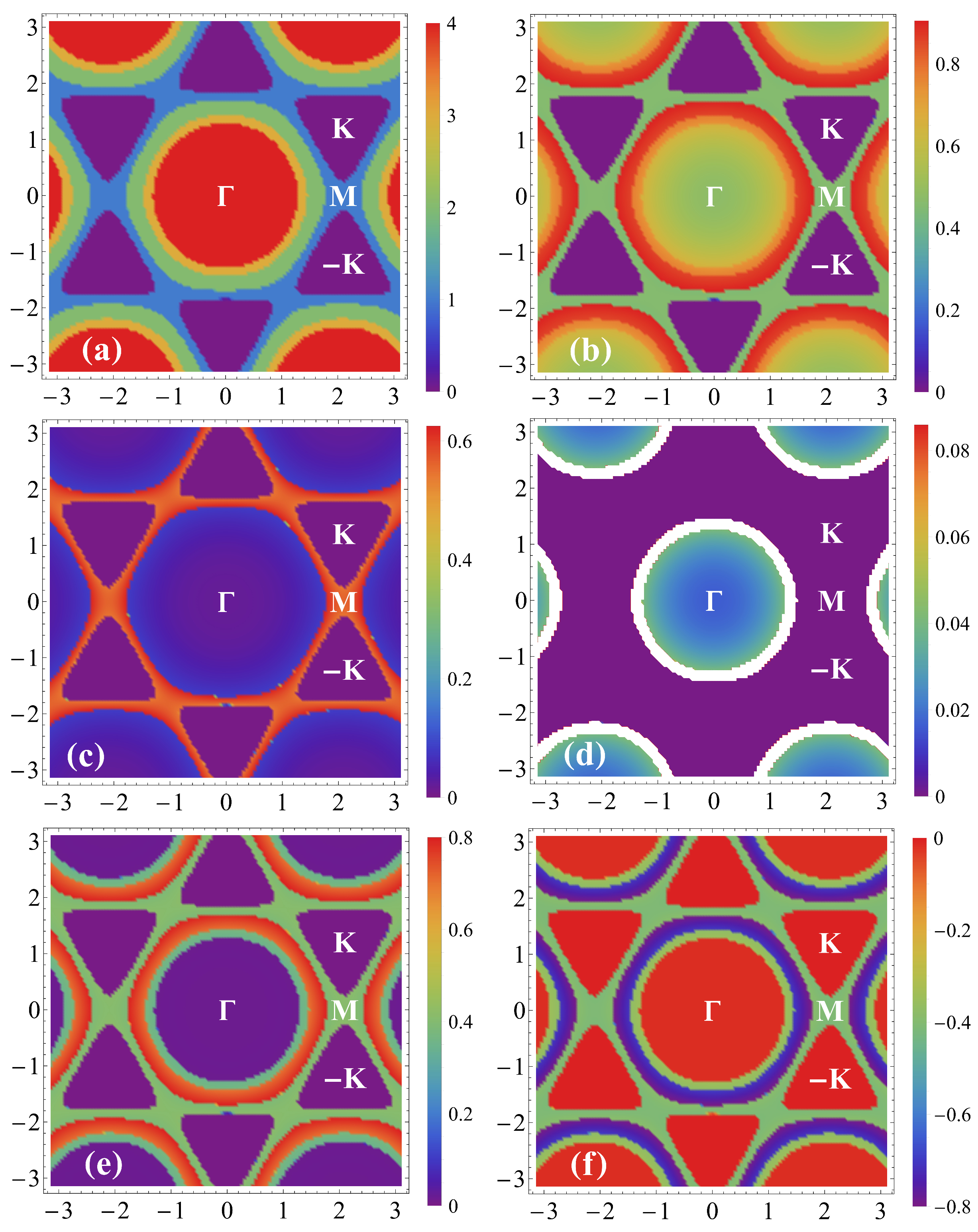}
\caption{\label{fig:Fig_7} The ${\bf{k}}$-map for the calculated key physical quantities for doped AA bilayer graphene with $x=1$. Specifically, the following results are illustrated: panel (a): chemical potential $\mu$, panel (b): antiferromagnetic order parameter $\Delta_{\rm AFM}$, panel (c): excitonic order parameter $\Delta_{\uparrow}$, panel (d) excitonic order parameter $\Delta_{\downarrow}$, panel (e): average charge density imbalance function $\delta \bar{n}_{1}$, and panel (f): average charge density imbalance function $\delta \bar{n}_{2}$. The intralayer Coulomb potential is fixed at $U = 3.3\gamma_0$, while the magnetic field parameter $\tilde{B}$ is set to $\tilde{B} = 0.1$. Additionally, the layer voltages are set to $V_1 = -0.5 \gamma_0=-1.4$ \text{eV} and $V_2 = 0.5 \gamma_0=1.4$ \text{eV}.}
\end{figure}

\begin{figure}[h!]
\includegraphics[scale=0.2]{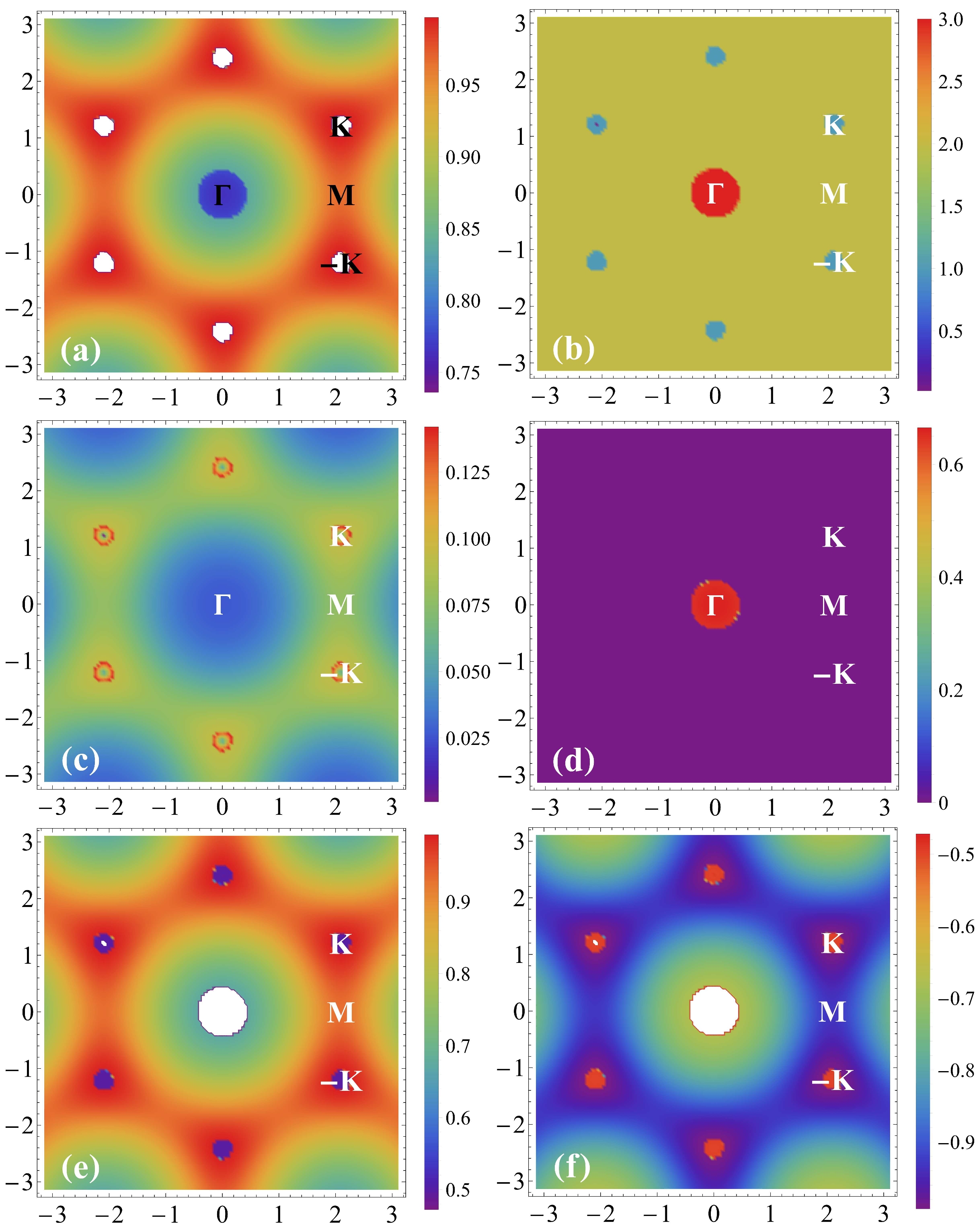}
\caption{\label{fig:Fig_8} The ${\bf{k}}$-map for the calculated key physical quantities for doped AA bilayer graphene with $x=1$. Specifically, the following results are illustrated: panel (a): chemical potential $\mu$, panel (b): antiferromagnetic order parameter $\Delta_{\rm AFM}$, panel (c): excitonic order parameter $\Delta_{\uparrow}$, panel (d) excitonic order parameter $\Delta_{\downarrow}$, panel (e): average charge density imbalance function $\delta \bar{n}_{1}$, and panel (f): average charge density imbalance function $\delta \bar{n}_{2}$. The intralayer Coulomb potential is fixed at $U = 3.3\gamma_0$, while the magnetic field parameter $\tilde{B}$ is set to $\tilde{B} = 0.5$. Additionally, the layer voltages are set to $V_1 = -0.5 \gamma_0=-1.4$ \text{eV} and $V_2 = 0.5 \gamma_0=1.4$ \text{eV}.}
\end{figure}

The implications of these findings are significant, with potential applications in modern quantum transport and spintronic devices. Additionally, this research may influence the development of new types of spin-valve devices based on AA-stacked bilayer graphene (BLG). Understanding the dynamics of spin-controlled excitonic states could pave the way for innovative spin-controlled quantum devices and heterostructures, potentially enhancing device performance and functionality in future technological applications. 
At the end of this section, we present the $\mathbf{k}$-distributions of important calculated physical quantities in Figs. \ref{fig:Fig_7} and \ref{fig:Fig_8}. These distributions are analyzed for two different values of the external magnetic field: $\tilde{B} = 0.1$ (shown in Fig. \ref{fig:Fig_7}), and $\tilde{B} = 0.5$ (shown in Fig. \ref{fig:Fig_8}).
These figures provide a visual representation of how the physical quantities vary with the momentum space under different magnetic field conditions.
%
%
We observe that for the higher value of the magnetic field (see in Fig. \ref{fig:Fig_8}, where $B=\mu_{\rm B}B/\gamma_0=1.5$ Tesla), the behavior of the chemical potential (see panel (a) in Fig. \ref{fig:Fig_8}) is entirely opposite to that of the low magnetic field case (see panel (a) in Fig. \ref{fig:Fig_7}, where $B=0.3$ Tesla). In the higher magnetic field scenario, many ${\bf{K}}$-pockets around the Dirac points (illustrated in Fig. \ref{fig:Fig_8}) are fully occupied opposite to the case of smaller magnetic field , presented in Fig. \ref{fig:Fig_7}.
This contrasting behavior is evident also in the ${\bf{k}}$-map of the excitonic gap parameter $\Delta_{\uparrow}$ (see panels (c), in Figs. \ref{fig:Fig_7} and \ref{fig:Fig_8}), as well as in the average charge density difference functions $\delta \bar{n}_{1}$ (see panels (e) in Figs. \ref{fig:Fig_7} and \ref{fig:Fig_8}).
At $\tilde{B} = 0.5$, the ${\bf{k}}$-distributions of the antiferromagnetic order parameter $\Delta_{\rm AFM}\left({\bf{k}}\right)$ and the excitonic gap parameter $\Delta_{\downarrow}\left({\bf{k}}\right)$ reveal a single peak at the symmetry point $\Gamma$ in reciprocal space (see panels (b) and (d) in Fig. \ref{fig:Fig_8}). 
%
\subsection{\label{sec:Section_3_3} Doping dependence}
\begin{figure}[h!]
\includegraphics[scale=0.31]{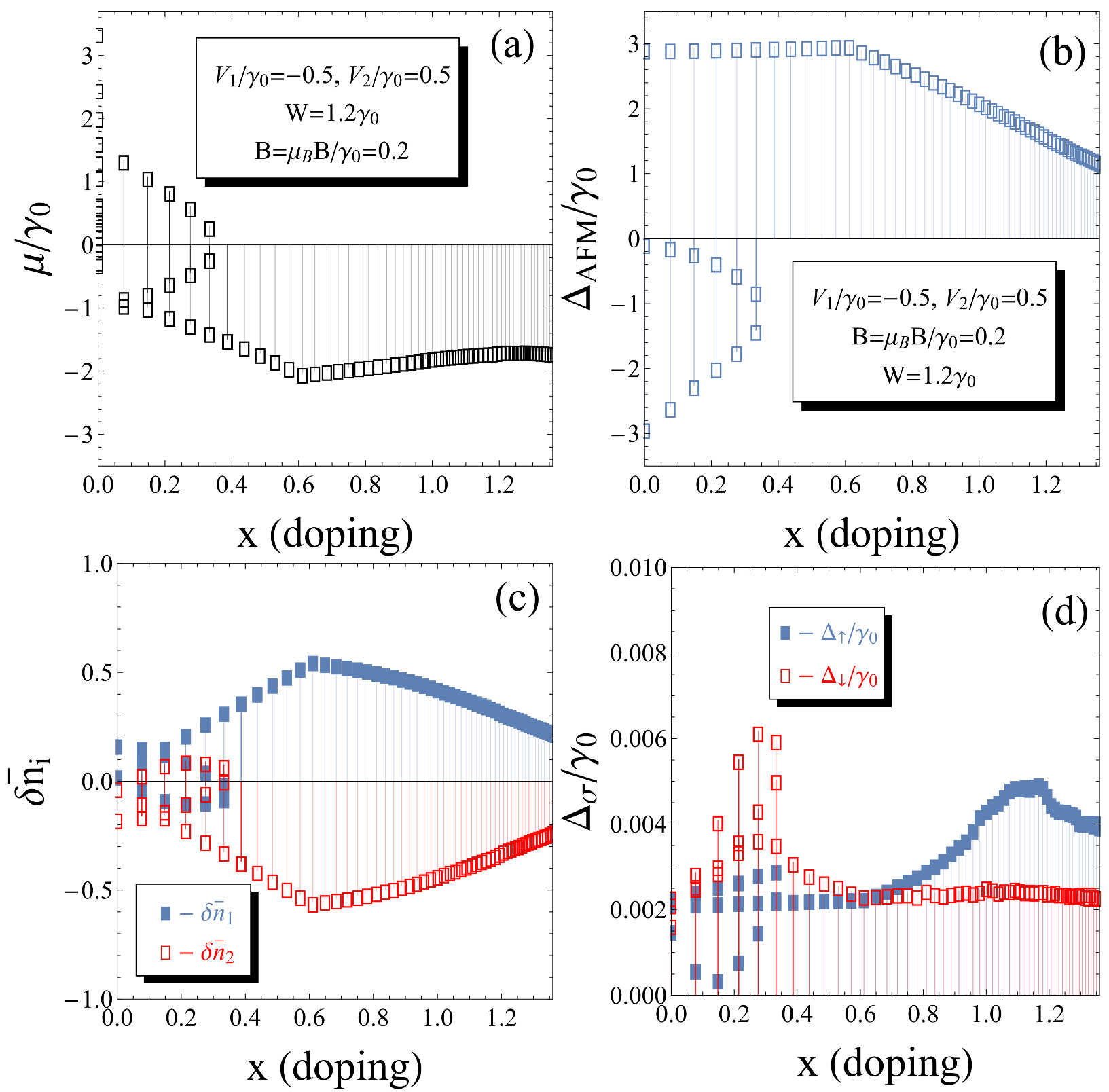}
\caption{\label{fig:Fig_9} The numerical solutions for key physical quantities are presented as functions of the doping parameter $x$, defined in Eq.(\ref{Equation_24}). Specifically, the following results are illustrated: panel (a): chemical potential $\mu$, panel (b): antiferromagnetic order parameter $\Delta_{\rm AFM}$, panel (c): average density imbalance functions $\delta \bar{n}_{1}$ and $\delta \bar{n}_{2}$, panel (d): excitonic order parameters $\Delta_{\uparrow}$ and $\Delta_{\downarrow}$. The intralayer Coulomb potential is fixed at $U = 3.3\gamma_0$, and the interlayer Coulomb potential at $W=1.2\gamma_0$. The magnetic field parameter $\tilde{B}$ is set to $\tilde{B} = 0.2$. Additionally, the layer voltages are set to $V_1 = -0.5 \gamma_0=-1.4$ \text{eV} and $V_2 = 0.5 \gamma_0=1.4$\text{eV}.}
\end{figure}

\begin{figure}[h!]
\includegraphics[scale=0.3]{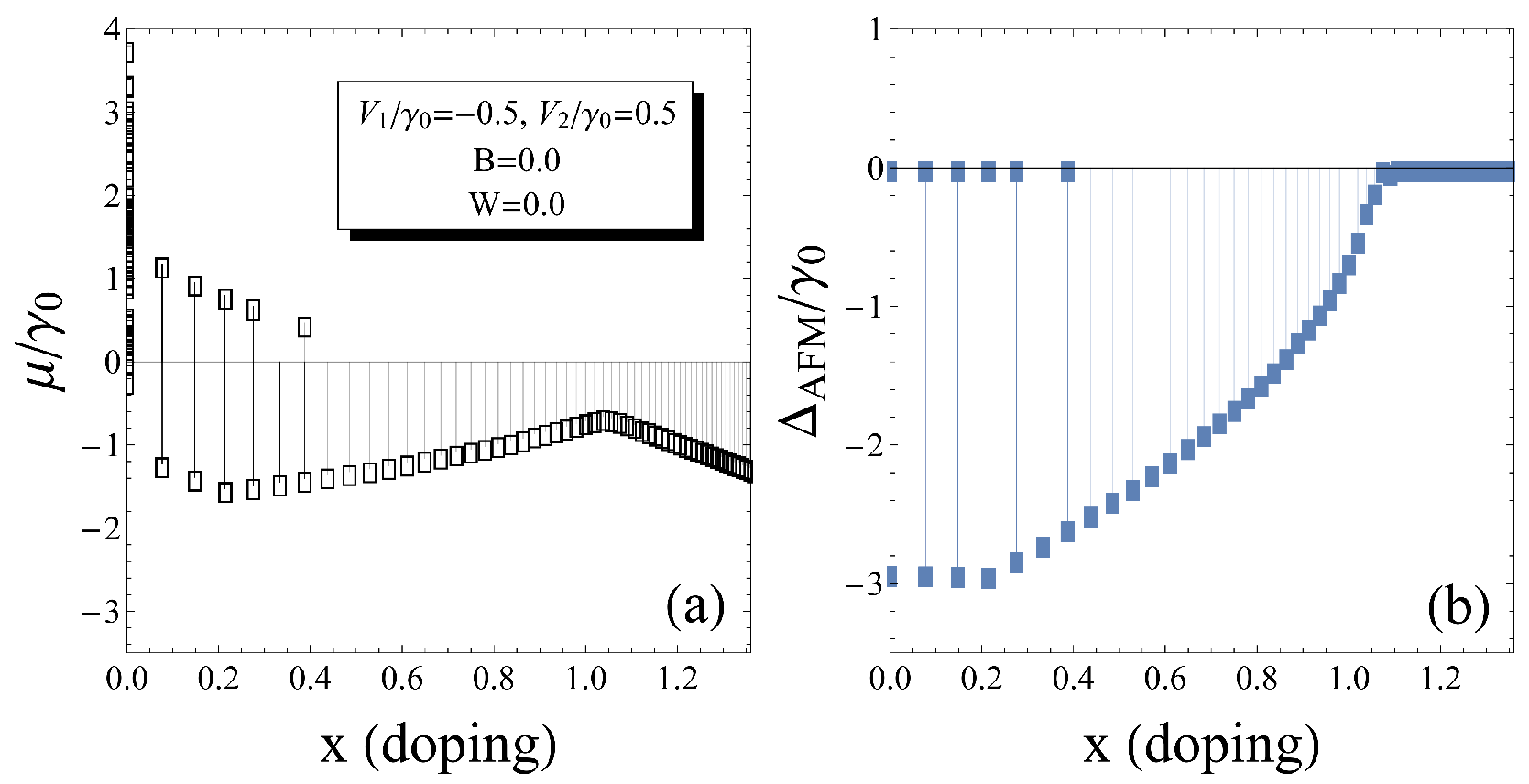}
\caption{\label{fig:Fig_10} The numerical solutions for key physical quantities are presented as functions of the doping parameter $x$, defined in Eq.(\ref{Equation_24}). Specifically, the following results are illustrated: panel (a): chemical potential $\mu$, panel (b): antiferromagnetic order parameter $\Delta_{\rm AFM}$. The intralayer Coulomb potential is fixed at $U = 3.3\gamma_0$, and the interlayer Coulomb potential at $W=0$. The magnetic field parameter $\tilde{B}$ is set to $\tilde{B} = 0$. Additionally, the layer voltages are set to $V_1 = -0.5 \gamma_0=-1.4$ \text{eV} and $V_2 = 0.5 \gamma_0=1.4$ \text{eV}.}
\end{figure}

\begin{figure}[h!]
\includegraphics[scale=0.55]{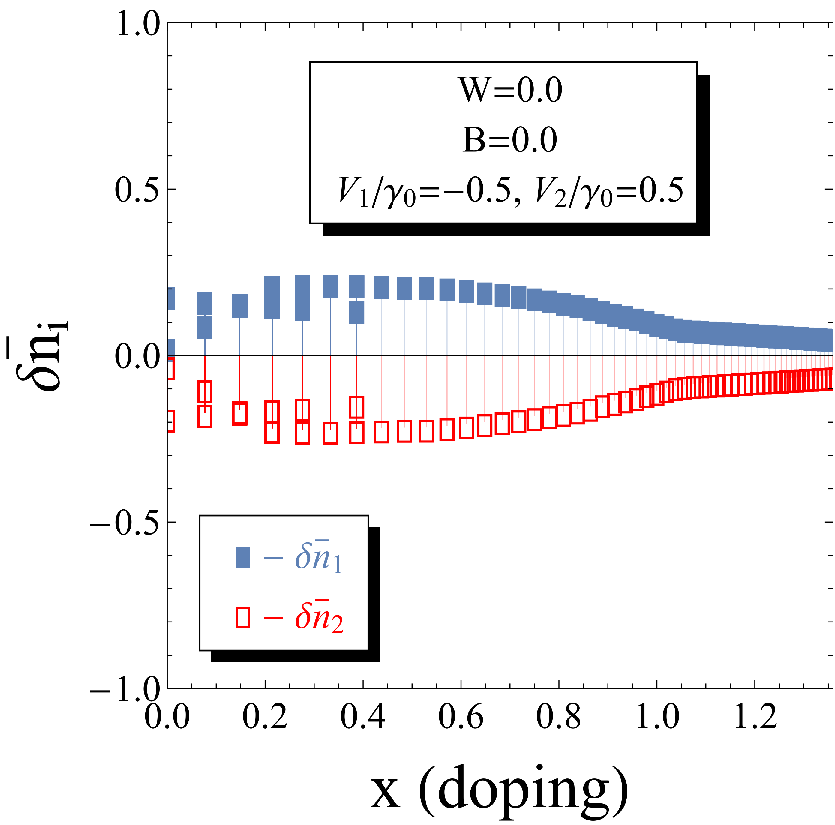}
\caption{\label{fig:Fig_11} The numerical solutions for average density imbalance functions $\delta \bar{n}_{1}$ and $\delta \bar{n}_{2}$ as functions of the doping parameter $x$, defined in Eq.(\ref{Equation_24}). The intralayer Coulomb potential is fixed at $U = 3.3\gamma_0$, and the interlayer Coulomb potential at $W=1.2\gamma_0$. The magnetic field parameter $\tilde{B}$ is set to high value $\tilde{B} = 0.8$. Additionally, the electric field potentials are set to $V_1 = -0.5 \gamma_0=-1.4$ \text{eV} and $V_2 = 0.5 \gamma_0=1.4$ \text{eV}.}
\end{figure}

\begin{figure}[h!]
\includegraphics[scale=0.31]{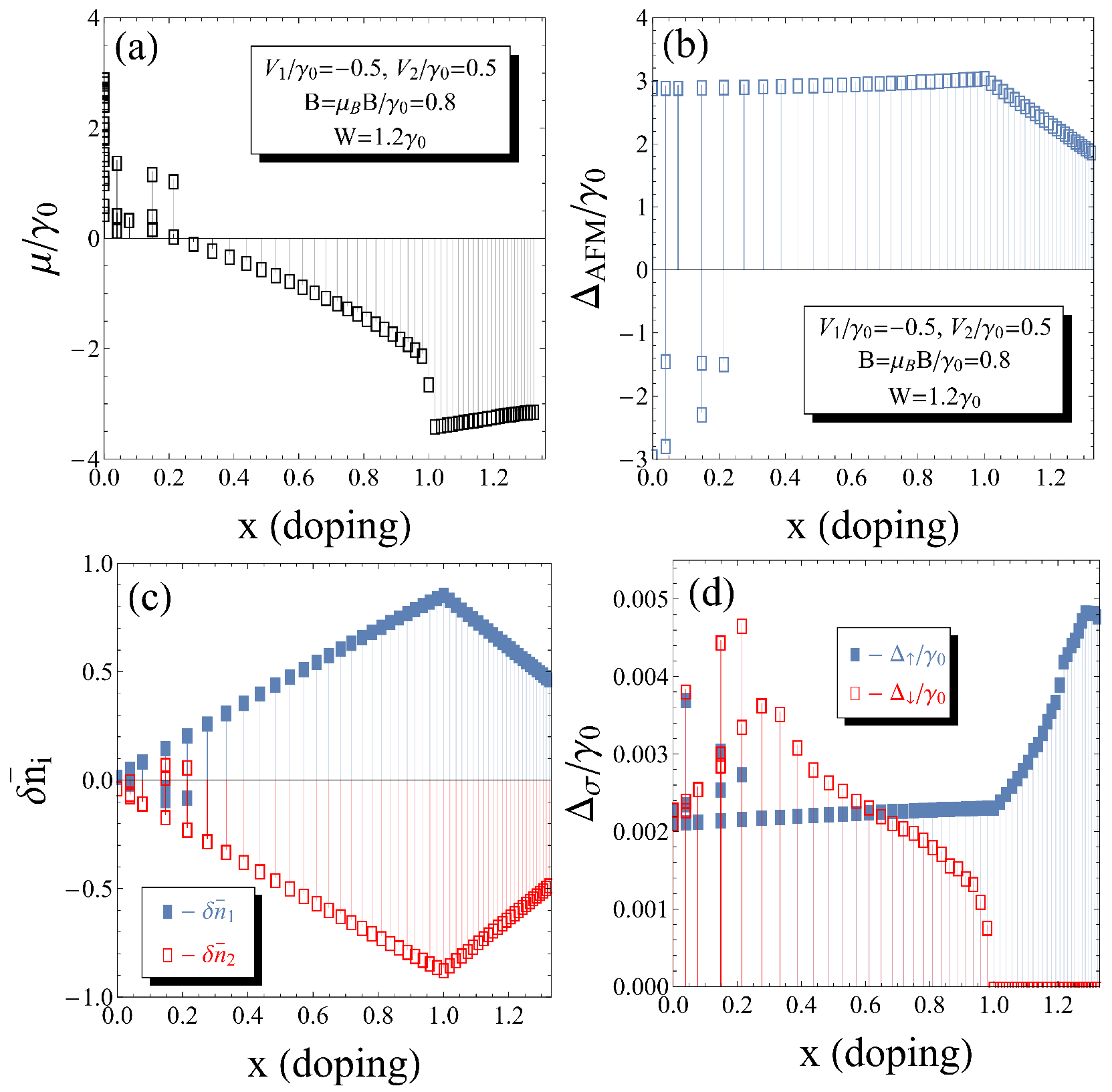}
\caption{\label{fig:Fig_12} The numerical solutions for key physical quantities are presented as functions of the doping parameter $x$, defined in Eq.(\ref{Equation_24}). Specifically, the following results are illustrated: panel (a): chemical potential $\mu$, panel (b): antiferromagnetic order parameter $\Delta_{\rm AFM}$, panel (c): average density imbalance functions $\delta \bar{n}_{1}$ and $\delta \bar{n}_{2}$, panel (d): excitonic order parameters $\Delta_{\uparrow}$ and $\Delta_{\downarrow}$. The intralayer Coulomb potential is fixed at $U = 3.3\gamma_0$, and the interlayer Coulomb potential at $W=1.2\gamma_0$. The magnetic field parameter $\tilde{B}$ is set to $\tilde{B} = 0.2$. Additionally, the layer potentials are set to $V_1 = -0.5 \gamma_0=-1.4$ \text{eV} and $V_2 = 0.5 \gamma_0=1.4$\text{eV}.}
\end{figure}

%
In this section, we discuss the doping dependence of the calculated physical parameters. The doping is defined in Eq. (\ref{Equation_24}) in Section \ref{sec:Section_2_2}. Figures \ref{fig:Fig_9}, \ref{fig:Fig_10},\ref{fig:Fig_11} and \ref{fig:Fig_12} present our findings based on different values for the magnetic field parameter $\mu_{\rm B}B/\gamma_0$ and the interlayer Coulomb interaction parameter $W$.

Specifically, we consider the following cases:

\begin{itemize}
	\item 
$\mu_{\rm B}B/\gamma_0 = 0.2$ (low magnetic field limit), presented in Fig. \ref{fig:Fig_9}.
\item 
$\mu_{\rm B}B/\gamma_0 = 0$ (zero magnetic field), shown in Figs. \ref{fig:Fig_10} and \ref{fig:Fig_11}.
\item 
$\mu_{\rm B}B/\gamma_0 = 0.8$ (high magnetic field limit), depicted in Fig. \ref{fig:Fig_12}.
\end{itemize}
For the calculations, the intralayer Coulomb potential is fixed at $U = 3.3 \gamma_0$, while the interlayer potential is set to $W = 1.2 \gamma_0$ (as seen in Figs. \ref{fig:Fig_9} and \ref{fig:Fig_12}), and $W = 0$ (as shown in Figs. \ref{fig:Fig_10} and \ref{fig:Fig_11}). The layer voltages are consistently set to $V_1 = -0.5 \gamma_0 = -1.4$, \text{eV} and $V_2 = 0.5 \gamma_0 = 1.4$, \text{eV}. All calculations were performed at a temperature of $T = 0$.

We observe that all parameters exhibit multivalued behavior in the limit of low doping, particularly as they approach the half-filling point. Notably, in panels (b) of Figs. \ref{fig:Fig_9}, \ref{fig:Fig_10} and \ref{fig:Fig_12}, at $x=0$ (the half-filling case), the overall antiferromagnetic order in the system is zero. However, as doping is increased, the system transitions into an antiferromagnetic ground state: for $\tilde{B} = 0.2$, we find that $\Delta_{\rm AFM} > 0$ with a value of $\Delta_{\rm AFM}=2.95 \gamma_0$ starting from $x = 0.38$ (see panel (b) in Fig.\ref{fig:Fig_9}); and for $\tilde{B} = 0.8$, $\Delta_{\rm AFM} > 0$ with a value of $\Delta_{\rm AFM}=2.93 \gamma_0$ starting from $x = 0.21$ (see panel (b) in Fig. \ref{fig:Fig_12}).

In contrast, when $\tilde{B} = 0$ (see in panel (b), in Fig.\ref{fig:Fig_10}), we observe a purely negative antiferromagnetic order parameter $\Delta_{\rm AFM} < 0$ beginning at $x = 0.33$ (with a value of $\Delta_{\rm AFM}=-2.71 \gamma_0$). This antiferromagnetic order is vanishing, furthermore, at the very high doping value $x=1.107$. Thus, by applying the magnetic field, the antiferromagnetic order transitions from negative to positive as the doping reaches a certain threshold value.
%
When comparing the average charge density differences $\delta \bar{n}_{i}$ in panels (c) of Figs.\ref{fig:Fig_9}, \ref{fig:Fig_11}, and in Fig. \ref{fig:Fig_12}, we observe that the values of $\delta \bar{n}_{1}$ are relatively small when $\tilde{B} = 0$. Additionally, we find that $\delta \bar{n}_{1}(\tilde{B} = 0.2) < \delta \bar{n}_{1}(\tilde{B} = 0.8)$. This indicates that increasing the magnetic field results in an enhancement of $\delta \bar{n}_{1}$.

Conversely, for the second layer, we note that $\delta \bar{n}_{2}(\tilde{B} = 0.2) > \delta \bar{n}_{2}(\tilde{B} = 0.8)$, suggesting that the magnetic field has the opposite effect on $\delta \bar{n}_{2}$ compared to $\delta \bar{n}_{1}$.

Furthermore, we observe, in Figs. \ref{fig:Fig_9} and \ref{fig:Fig_12}, a notable trend regarding the values of the excitonic order parameters $\Delta_{\uparrow}$ and $\Delta_{\downarrow}$ across different limits of electron doping. Specifically, as illustrated in panels (d) in Figs. \ref{fig:Fig_9} and \ref{fig:Fig_12}, we find that $\Delta_{\downarrow} > \Delta_{\uparrow}$ in the low doping limit, while the situation reverses in the case of large doping, leading to $\Delta_{\uparrow} > \Delta_{\downarrow}$.
%
%
At the end of this section, we should mentioned that the magnetic field and doping controlled formations of the antiferromagnetic and excitonic states, as well as the calculated average charge densities in the layers, could be verified experimentally using a combination of magnetic spin susceptibility and angle resolved exciton photoluminiscence measurements, as provided in Ref. \cite{cite_59}.  

\subsection{\label{sec:Section_3_4} Electronic band structure}
%
In this section, we calculate the electronic band structure corresponding to spin-$\uparrow$ and spin-$\downarrow$ directions using the formulae given in Eqs.(\ref{Equation_30}) and (\ref{Equation_34}) from Section \ref{sec:Section_2_3}. The evaluations of the band structure are conducted for a variety of parameters in the system, leveraging the numerical solutions of physical quantities presented in the previous section.

\begin{figure}[h!]
\includegraphics[scale=0.53]{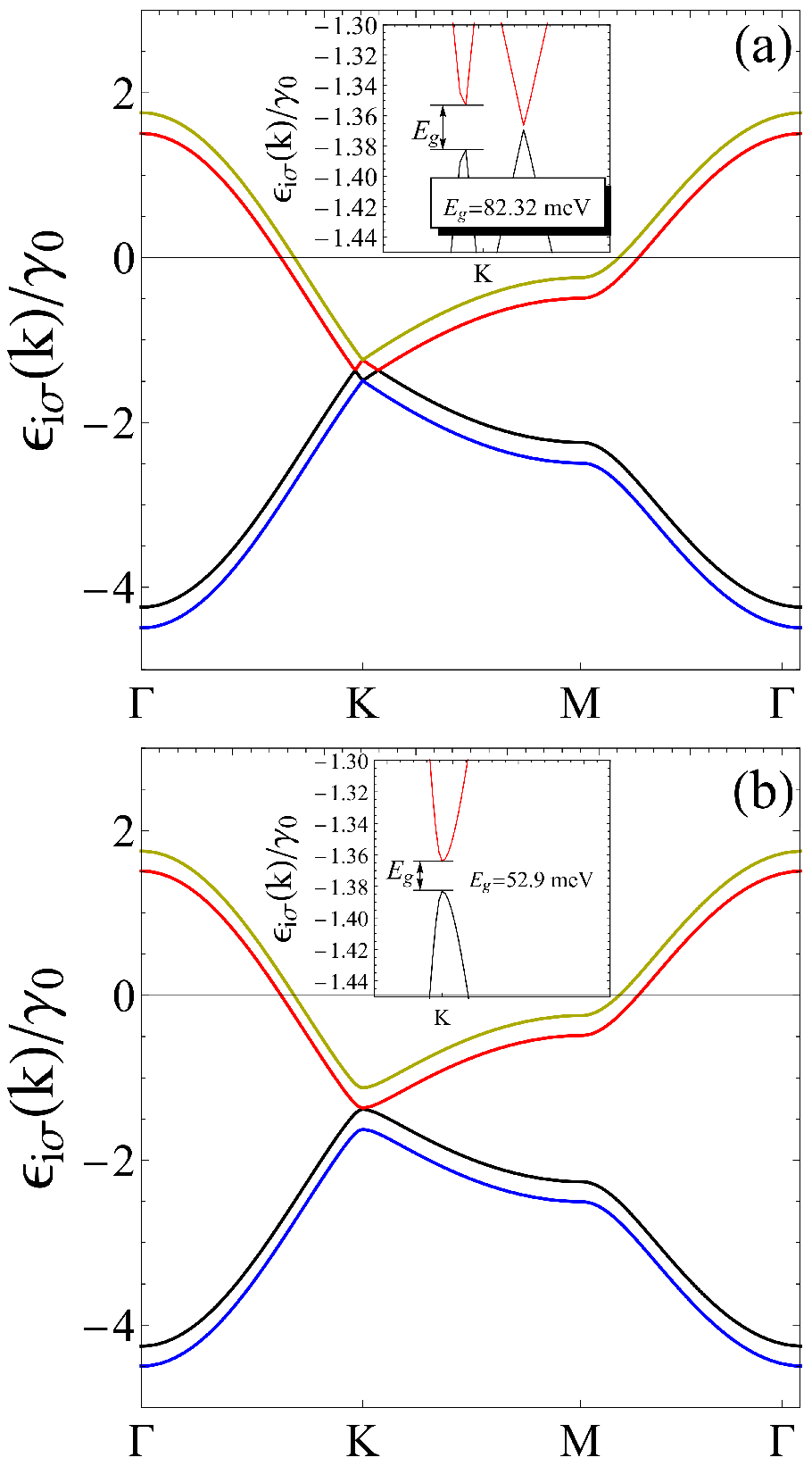}
\caption{\label{fig:Fig_13} The electronic band structure in the doped (with $x=1.0$) double gated AA BLG system in the regimes with $W=1.2\gamma_0$, $V_{1}=0$ and $V_{2}=0$ (see in panel (a)) and $W=0.0$, $V_{1}=0.2\gamma_0=0.56$ \text{eV} and $V_{2}=0.5\gamma_0=1.4$ \text{eV} (see in panel (b)). The calculations are evaluated at zero magnetic field limit $\tilde{B}=0$. The intralayer Coulomb interaction parameter was fixed at value $U=3.3\gamma_0$ and temperatures was set to $T=0$.}
\end{figure}

Panels (a) and (b) of Fig. \ref{fig:Fig_13} illustrate the electronic band structure under the effects of the interlayer Coulomb interaction and the applied layer voltages, respectively, in the zero magnetic field limit ($\tilde{B} = 0$) and in the doped case ($x=1.0$). Specifically, panel (a) showcases the influence of the interlayer Coulomb interaction, while panel (b) focuses on the impact of the applied layer voltages.

The values of other physical quantities in Eqs. (\ref{Equation_30}) and (\ref{Equation_34}), such as the chemical potential, antiferromagnetic order parameter, excitonic gap parameters, and average charge density differences, have been derived from the solution of the self-consistent equations outlined in Eq.(\ref{Equation_36}) in Section \ref{sec:Section_2_3}.   
%
In the case where the electric field potentials are zero (refer to panel (a) in Fig. \ref{fig:Fig_13}), we used the following solutions as inputs for the electronic band structure calculation: chemical potential: $\mu = -0.542 \gamma_0 = -1.517$ \text{eV},
antiferromagnetic order parameter: $\Delta_{\rm AFM} = 0$, excitonic order parameters: $\Delta_{\uparrow} = 0.00442 \gamma_0 = 12.37$ \text{meV} and $\Delta_{\downarrow} = 0.00442 = 12.37$ \text{meV}, average charge density differences: $\delta \bar{n}_{1} = \delta \bar{n}_{2} = 0$.
As shown in panel (a) of Fig. \ref{fig:Fig_13}, a sufficiently large bandgap exists near the Dirac point between the quasienergies $\varepsilon_{1{\bf{k}}\sigma}$ and $\varepsilon_{2{\bf{k}}\sigma}$, with a bandgap energy of $E_{\rm g} = 82.32$ \text{meV}. The band structure is identical for both spin directions due to the doping level and the absence of antiferromagnetic order in the system.

In panel (b) of Fig. \ref{fig:Fig_13}, we calculate the electronic band structure under the condition of zero interlayer interaction (indicating the absence of excitons in the system). Here, we consider gated BLG with electric field potentials of $V_1 = 0.2 \gamma_0 = 0.56$ \text{eV} applied to the bottom layer and $V_2 = 0.5 \gamma_0 = 1.4$ \text{eV} applied to the top layer. The physical quantities used as inputs for this calculation are:chemical potential: $\mu = -0.548 \gamma_0 = -1.534 , \text{eV}$, antiferromagnetic order parameter: $\Delta_{\rm AFM} = 0$,
excitonic order parameters: $\Delta_{\uparrow} = 0$ and $\Delta_{\downarrow} = 0$, average charge density differences: $\delta \bar{n}_{1} = -0.036434$ and $\delta \bar{n}_{2} = -0.090979$.
In this scenario, we again identify the presence of a bandgap in the AA BLG system, with a value of $E_{\rm g} = 52.9$ \text{meV}, which is significantly smaller than the bandgap observed in the unbiased case (see in panel(a), in Fig.\ref{fig:Fig_13}). The insets in Fig. \ref{fig:Fig_13} illustrate the formation of bandgaps in the vicinity of the Dirac point for both discussed cases. 

Thus, the overarching conclusion from these calculations is that the electronic bandgap induced by the applied layer potentials is smaller compared to the bandgap resulting from excitonic effects in the unbiased AA BLG. Consequently, excitons play a crucial role in the doped AA BLG system, driving the formation of the bandgap.     
In the pictures what follow, we will discuss only the influence of the external electric field potentials on the electronic band structure (by setting $W=0$ and $\tilde{B}=0$), including the presence of the antiferromagnetic order at zero doping. 

\begin{figure}[h!]
\includegraphics[scale=0.53]{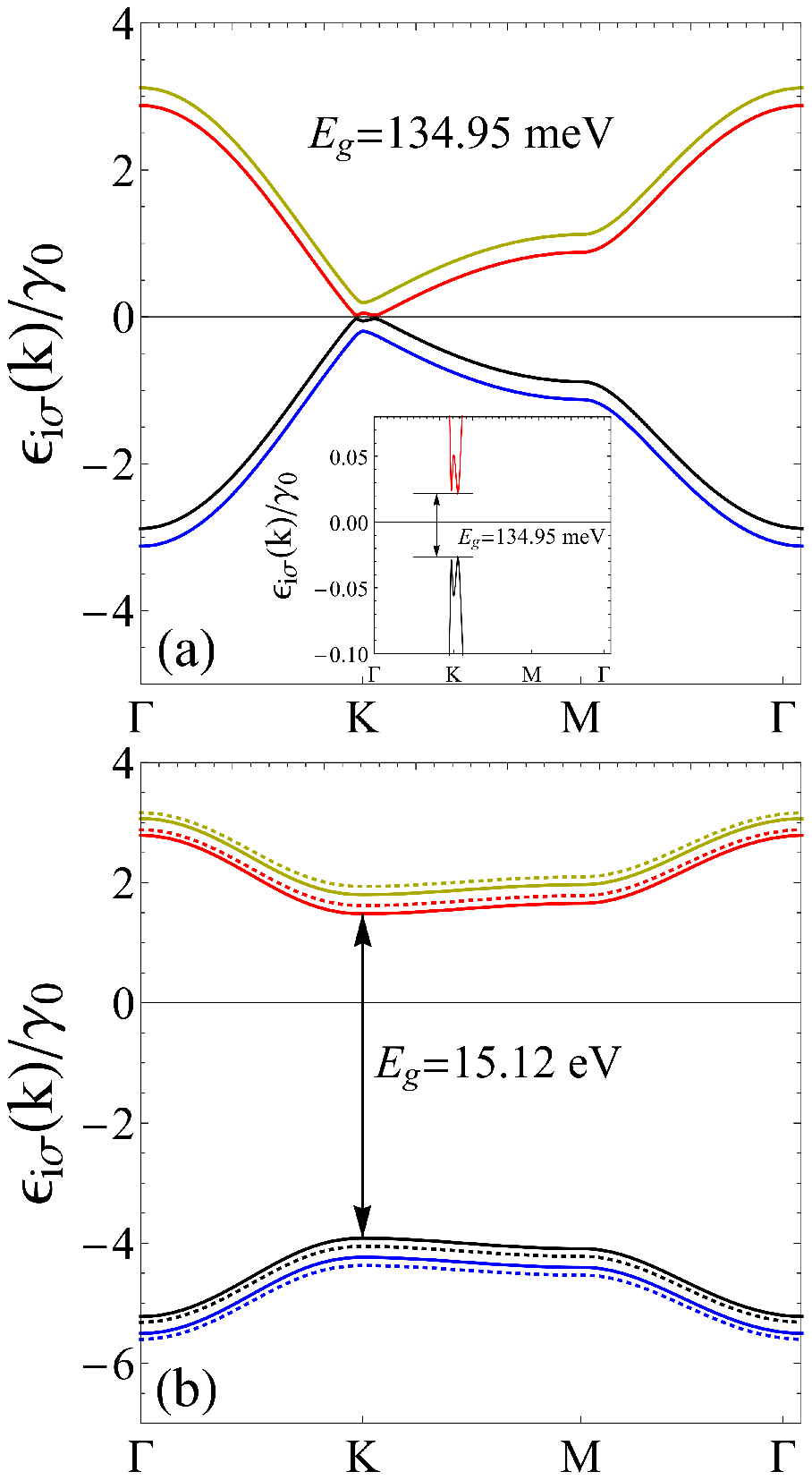}
\caption{\label{fig:Fig_14} The electronic band structure for the undoped ($x = 0$) double-gated AA-stacked bilayer graphene (BLG) system is analyzed under two different regimes. In panel (a), we consider the configuration with $W = 0$, $V_1 = 0.2\gamma_0 = 0.56$ \text{eV}, $V_2 = 0.5\gamma_0 = 1.4$\text{eV}, and the antiferromagnetic order parameter set to $\Delta_{\rm AFM} = 0$. In panel (b), we maintain the same conditions of $W = 0$, $V_1 = 0.2$, $\gamma_0 = 0.56$ \text{eV}, and $V_2 = 0.5\gamma_0 = 1.4$ \text{eV} but set the antiferromagnetic order parameter to lower bound solution $\Delta_{\rm AFM}=-\Delta_0$. All calculations are performed at zero magnetic field $\tilde{B} = 0$, with the intralayer Coulomb interaction parameter fixed at $U=3.3\gamma_0$.}
\end{figure}

In Fig. \ref{fig:Fig_14}, we present the calculated electronic band structure for the case of an undoped ($x = 0$) AA-stacked bilayer graphene (BLG) system at zero magnetic field and with the interlayer Coulomb interaction set to $W = 0$. The values of the external electric field potentials used in this calculation are the same as those applied in panel (b) in Fig. \ref{fig:Fig_13}.
%
The band structure presented in panel (a), in Fig.\ref{fig:Fig_14}, was evaluated for the zero branch of the antiferromagnetic order parameter solution (analogous to panel (b) in Fig. \ref{fig:Fig_4}) at zero magnetic field $\tilde{B} = 0$. The following solutions were used for this case:
chemical potential: $\mu_{\rm m} = 1.642 \gamma_0 = 4.597$ \text{eV}, antiferromagnetic order parameter: $\Delta_{\rm AFM} = 0$, average charge density differences: $\delta{\bar{n}}_{1} = -0.010764$ and $\delta{\bar{n}}_{2} = -0.027096$, excitonic order parameters: $\Delta_{\uparrow} = 0$ and $\Delta_{\downarrow} = 0$. In panel (a), we observe a "Mexican hat" shape in the band structure around the Dirac point, characterized by a bandgap on the order of $E_{g} = 134.95$, \text{meV}. Such a large band gap has been obtained recently in Ref.\cite{cite_13}, in the context of double-gated experimental setup for the bilayer graphene system. The large value of the bandgap $E_g$, in panel (a), indicates that the system is in a semiconducting-like state.

\begin{figure}[h!]
\includegraphics[scale=0.28]{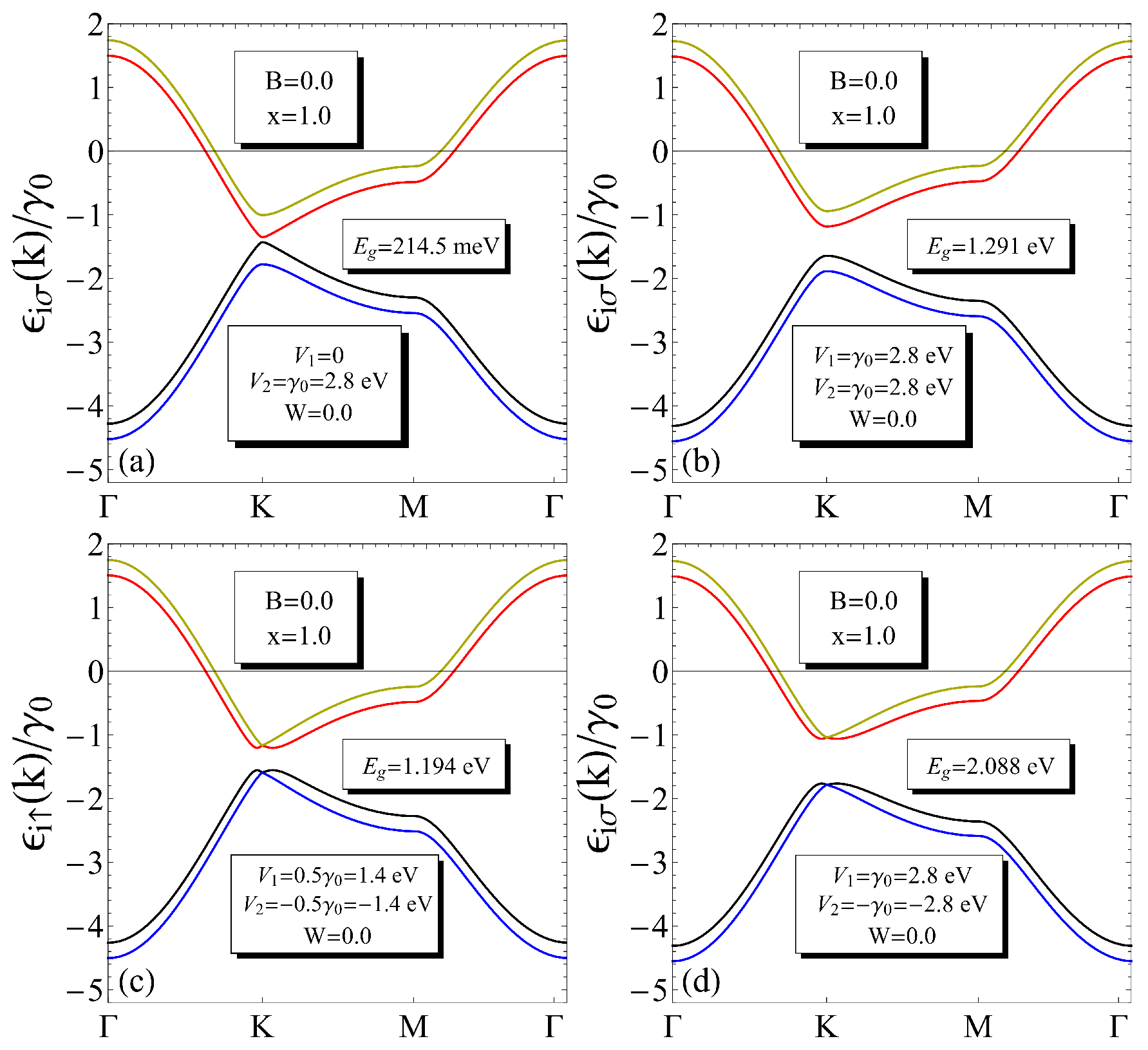}
\caption{\label{fig:Fig_15} The electronic band structure for the doped $(x = 1.0)$ double-gated AA-stacked bilayer graphene (BLG) system is examined under the following conditions: $W = 0$, $U = 3.3 \gamma_0$, $\tilde{B} = 0$. The values of the layer voltages employed in the calculations are as follows: panel (a): $V_{1} = 0$, $V_{2} = \gamma_0 = 2.8$ \text{eV}, panel (b): $V_{1} = \gamma_0=2.8$ \text{eV}, $V_{2} = \gamma_0 = 2.8$\text{eV}, panel (c): $V_{1} = 0.5\gamma_0=1.4$ \text{eV}, $V_{2} = -0.5\gamma_0=-1.4$ \text{eV}, panel (d): $V_{1} =\gamma_0= 2.8$ \text{eV}, $V_{2} = -\gamma_0=-2.8$ \text{eV}.}
\end{figure}

Surprisingly, a significantly larger bandgap is obtained in panel (b) in Fig.\ref{fig:Fig_14}, where the band structure is calculated for the negative branch of the antiferromagnetic order parameter at zero doping (see in panel (b), in Fig.\ref{fig:Fig_4}). The following solutions were applied in this scenario: chemical potential: $\mu_{\rm m} = 0.423 \gamma_0 = 1.184$ \text{eV}, antiferromagnetic order parameter: $\Delta_{\rm AFM} = -\Delta_{0} = -2.925 \gamma_0 = -8.19$ \text{eV}, average charge density differences: $\delta{\bar{n}}_{1} = -0.010764$ and $\delta{\bar{n}}_{2} = -0.027096$, excitonic order parameters: $\Delta_{\uparrow} = 0$ and $\Delta_{\downarrow} = 0$. The chemical potential $\mu_{\rm m}$ used in this context represents the average value of the chemical potential taken from the multiple solutions at $x=0$ similar to what is shown in panel (a) of Fig. \ref{fig:Fig_4}). In this instance, the obtained bandgap is on the order of $E_{g} = 15.12$, \text{eV}, suggesting that the AA BLG system is in an insulating state under these conditions.     
%
The electronic band structure, for the doped AA BLG case (with $x=1.0$) has been evaluated in Fig.\ref{fig:Fig_15} for a range of values of the applied layer voltages. These configurations allow us to analyze how the variations in layer voltages affect the electronic band structure of the doped AA BLG system. These calculations are performed under the following conditions: $W = 0$, $U = 3.3 \gamma_0$, $\tilde{B} = 0$, and $T = 0$.
%
In panel (a), we observe that a bandgap of approximately $E_g = 214.5$ \text{meV} opens up for the case where $V_{1} = 0$ and $V_{2} = \gamma_0 = 2.8$ \text{eV}. This indicates that the system is in a semiconducting-like state. The physical quantities used in this calculation are derived from the solutions of the self-consistent equations in Eq.(\ref{Equation_36}), with the following values: chemical potential: $\mu = -0.564 \gamma_0 = -1.579$ \text{eV}, antiferromagnetic order parameter: $\Delta_{\rm AFM} = 0$ \text{eV}, average charge density differences: $\delta{\bar{n}}_{1} = -0.000622$ and $\delta{\bar{n}}_{2} = -0.182904$, excitonic order parameters: $\Delta_{\uparrow} = 0$ and $\Delta_{\downarrow} = 0$. 

In panel (b), we calculate the band structure with increased layer potential $V_{1} = 2.8$\text{eV} while keeping $V_{2} = \gamma_0 = 2.8$ \text{eV}. In this scenario, the system exhibits insulating behavior, with the following parameter values: chemical potential: $\mu = -0.587 \gamma_0 = -1.643$ \text{eV}, antiferromagnetic order parameter: $\Delta_{\rm AFM} = 0$ \text{eV}, average charge density differences: $\delta{\bar{n}}_{1} = -0.179789$ and $\delta{\bar{n}}_{2} = -0.179789$, excitonic order parameters: $\Delta_{\uparrow} = 0$ and $\Delta_{\downarrow} = 0$.
The observed bandgap is significantly larger, with $E_g = 1.291$ \text{eV}. In panel (c), when we change the layer voltages to $V_{1} = 0.5 \gamma_0 = 1.4$ \text{eV} and $V_{2} = -0.5 \gamma_0 = -1.4$ \text{eV}, the system remains in an insulating state, and we observe a "Mexican hat" structure near the Dirac point with a large bandgap of $E_g = 1.194$ \text{eV}. The parameters used here include: chemical potential: $\mu = -0.553 \gamma_0 = -1.548$ \text{eV}, antiferromagnetic order parameter: $\Delta_{\rm AFM} = 0$ \text{eV}, average charge density differences: $\delta{\bar{n}}_{1} = -0.090147$ and $\delta{\bar{n}}_{2} = 0.090147$, excitonic order parameters: $\Delta_{\uparrow} = 0$ and $\Delta_{\downarrow} = 0$.
Furthermore, in panel (d), we used $\mu=-0.586\gamma_0=-1.643$ \text{eV}, $\Delta_{\rm AFM}=0$ \text{eV}, $\delta{\bar{n}}_{1}=-0.178555$, $\delta{\bar{n}}_{2}=0.178555$, $\Delta_{\uparrow}=0$ and $\Delta_{\downarrow}=0$. In this configuration, we achieve a bandgap of approximately $E_g = 2.088$ \text{eV}, which is nearly twice as large as the bandgap observed in panel (c).
The overall conclusion drawn from the calculations presented in Figs. \ref{fig:Fig_13} - \ref{fig:Fig_15} is that varying the layer potentials in the AA BLG structure allows for the exploration of different states within the system. Specifically, we can observe the following:
non-gapped AA BLG: exhibiting metallic behavior, excitonic insulator state: as evidenced in Fig. \ref{fig:Fig_13}, semiconducting AA BLG: displayed in Figs. \ref{fig:Fig_14} and \ref{fig:Fig_15}, strong insulator state: also indicated in Figs. \ref{fig:Fig_14} and \ref{fig:Fig_15}.
These findings highlight the tunability of the electronic properties of the AA-stacked bilayer graphene system through external electric field potentials, offering insights into potential applications in electronic and optoelectronic devices.
%
\section{\label{sec:Section_4} Conclusion}
%
We have considered, for the first time, the double-gated structure of AA-stacked bilayer graphene, where layer voltages are applied separately to the sublattices in both layers. Additionally, we examined the effects of a magnetic field applied perpendicularly to the system and discussed the staggered antiferromagnetic order. Excitonic correlations in the system were analyzed, along with various competing orders that arise within this framework.

Our analytical calculations were conducted within the double-layer Hubbard model, incorporating the interlayer Coulomb interaction. Under this model, the average particle occupancy in the upper layer becomes significantly dependent on the on-site particle occupancy in the bottom layer and on the interaction potentials present in the system. 

We demonstrated that the behavior of the calculated physical quantities is strongly influenced by the doping levels in the system as well as the applied external electric field potentials $V_1$ and $V_2$. Important quantities calculated within this work include the chemical potential in the AA BLG, the antiferromagnetic order parameter, average particle occupations at the sublattice sites, and excitonic order parameters corresponding to different spin directions. We estimated the order of magnitude of the intralayer on-site Coulomb interaction parameter and highlighted its localizing effect on particles at the sublattice site positions. The results are consistent with existing findings in the literature.

We also showed how doping significantly alters the solutions of the obtained self-consistent equations: in the undoped case, we found multivalued solutions, while at certain doping levels, the solutions became single-valued. Analyzing the calculated physical quantities as functions of the magnetic field, we identified a critical value of the magnetic field parameter at which these quantities undergo substantial changes. Notably, we observed a Wigner-type localization in the average charge density differences between sublattice sites, with the excitonic gap parameter closing above the critical magnetic field value, while another remains open and constant in the high magnetic field limit. Moreover, we investigated the doping dependence of the excitonic order parameters, observing that the excitonic gap in one spin channel closes at a critical doping level, whereas it increases in another channel. This finding suggests potential pathways for spin-controlled electronic transport in such systems and may facilitate the development of spin-valve effects.

Furthermore, we calculated the electronic band structure in the AA BLG system and observed unusual behavior not typically documented in the literature. Specifically, instead of the expected linearly crossing bands with zero gap, we found a gapped band structure at the Dirac point when varying the applied layer voltages. We attribute this bandgap to the geometry of the applied layer voltages, revealing a series of states that can be manipulated by adjusting $V_1$ and $V_2$. The system exhibits excitonic insulator bandgap formation in the unbiased state, transitioning into an excitonic insulator phase. Moreover, the AA BLG system demonstrates semiconducting and insulating behaviors depending on the signs and magnitudes of the applied potentials and the doping levels.

The possible reasonable realization of presented results in the paper could be done via the pulse electric field experimental techniques, as presented in Refs.\cite{cite_60, cite_61}, in a good connection with graphene and graphene based materials. Indeed, the Wannier-Stark localization, which is observed in the mentioned references will provide a relatively smooth perturbation of the electronic system when providing large voltages.   

The results obtained in this study are significant from both a theoretical and experimental perspective, representing a substantial contribution to the research on graphene-based materials. They suggest significant potential applications for the AA BLG system as a large bandgap semiconducting material and open new avenues for the development of spin-valve devices.
%
\section{\label{sec:Section_5} The calculation of important averages and coefficients}
%
Here, we will calculate the coefficients $\alpha^{\sigma}_{i{\bf{k}}}$, $\beta^{\sigma}_{ji{\bf{k}}}$, $\gamma^{\sigma}_{i{\bf{k}}}$, and $\delta^{\sigma}_{i{\bf{k}}}$ (where $j=1,2$ and $i=1,...,4$), entering in the right-hand sides in Eq.(\ref{Equation_36}), discussed in the Section \ref{sec:Section_2_3}. The partition function of the system with the introduced source terms is 
\begin{eqnarray}
&&{\cal{Z}}=\int\left[{{\cal{D}}\bar{\Psi}_{\uparrow}{\cal{D}}\Psi}_{\uparrow}\right]\int\left[{{\cal{D}}\bar{\Psi}_{\downarrow}{\cal{D}}\Psi}_{\downarrow}\right]e^{-{\cal{S}}\left[\bar{\Psi},\Psi\right]}={\cal{Z}}_{\uparrow}{\cal{Z}}_{\downarrow}
\nonumber\\
&&=\int{\left[{\cal{D}}\bar{\Psi}_{\sigma}{\cal{D}}\Psi_{\sigma}\right]}e^{-\frac{1}{2}\sum_{{\bf{k}}\nu_{n},\sigma}\bar{\Psi}_{{\bf{k}}\sigma}\left(\nu_{n}\right){{\cal{G}}'}^{-1}_{{\bf{k}}\sigma}\left(\nu_{n}\right)\Psi_{{\bf{k}}\sigma}\left(\nu_{n}\right)}\times
\nonumber\\
&&\times{e^{\frac{1}{2}\sum_{{\bf{k}}\nu_{n},\sigma}\bar{\Psi}_{{\bf{k}}\sigma}\left(\nu_{n}\right)J_{{\bf{k}}\sigma}\left(\nu_{n}\right)+\frac{1}{2}\sum_{{\bf{k}}\nu_{n},\sigma}\bar{J}_{{\bf{k}}\sigma}\left(\nu_{n}\right){\Psi}_{{\bf{k}}\sigma}\left(\nu_{n}\right)}}
\nonumber\\
&&=e^{-\frac{1}{2}\sum_{{\bf{k}}\nu_{n},\sigma}\bar{\Psi}_{{\bf{k}}\sigma}\left(\nu_{n}\right){\cal{D}}_{{\bf{k}}\sigma}\left(\nu_{n}\right){\Psi}_{{\bf{k}}\sigma}\left(\nu_{n}\right)},
\label{Equation_A_1}
\end{eqnarray}
where ${{\cal{G}}'}^{-1}_{{\bf{k}}\sigma}\left(\nu_{n}\right)$, in Eq.(\ref{Equation_A_1}) is ${{\cal{G}}'}^{-1}_{{\bf{k}}\sigma}\left(\nu_{n}\right)=\left(2/\beta{N}\right){\cal{G}}^{-1}_{{\bf{k}}\sigma}\left(\nu_{n}\right)$ and ${\cal{D}}_{{\bf{k}}\sigma}\left(\nu_{n}\right)$ is the inverse matrix. We have introduced in Eq.(\ref{Equation_A_1}) the external fermionic source terms $J_{{\bf{k}}\sigma}\left(\nu_{n}\right)$ 
\begin{eqnarray} 
		J_{{\bf{k}}\sigma}(\nu_{n})=\left(
	\begin{array}{crrrr}
		J_{a_1{\bf{k}}\sigma}(\nu_{n})\\\\
		J_{b_1{\bf{k}}\sigma}(\nu_{n}) \\\\
		J_{a_2{\bf{k}}\sigma}(\nu_{n}) \\\\
		J_{b_2{\bf{k}}\sigma}(\nu_{n}) \\\\
	\end{array}
	\right),
	\label{Equation_A_2}
\end{eqnarray}
and its complex conjugate spinor $\bar{J}_{\sigma}\left({\bf{k}}, \nu_n\right)=\left(\bar{J}_{a1{\bf{k}}\sigma}\left(\nu_{n}\right),
\bar{J}_{b1{\bf{k}}\sigma}, \bar{J}_{a2{\bf{k}}\sigma},\bar{J}_{b2{\bf{k}}\sigma}\right)$.
We rewrite the system of equations in Eq.(\ref{Equation_36}) in the explicit form, in Fourier space by supposing that the excitonic pairing gap parameter is homogeneous and real: $\Delta^{\rm A}_{\sigma}=\bar{\Delta}_{\sigma}$ and $\Delta^{\rm A}_{\sigma}={\Delta^{\rm B}_{\sigma}}$. Therefore, we will write the self-consistent equations for the total excitonic gaps $\Delta^{\rm A}_{\sigma}+\bar{\Delta}^{\rm A}_{\sigma}+\Delta^{\rm B}_{\sigma}+\bar{\Delta}^{\rm B}_{\sigma}=4\Delta_{\sigma}$. We have the following complete set of the self-consistent equations: 

\begin{widetext}
\begin{eqnarray}
&&\frac{1}{\left(\beta{N}\right)^{2}}\sum_{{\bf{k}}\nu_{n},\sigma}\left(\left\langle \bar{a}_{1{\bf{k}}\sigma}\left(\nu_{n}\right)a_{1{\bf{k}}\sigma}\left(\nu_{n}\right)\right\rangle+\left\langle \bar{b}_{1{\bf{k}}\sigma}\left(\nu_{n}\right)b_{1{\bf{k}}\sigma}\left(\nu_{n}\right)\right\rangle+\left\langle \bar{a}_{2{\bf{k}}\sigma}\left(\nu_{n}\right)a_{2{\bf{k}}\sigma}\left(\nu_{n}\right)\right\rangle+\left\langle \bar{b}_{2{\bf{k}}\sigma}\left(\nu_{n}\right)b_{2{\bf{k}}\sigma}\left(\nu_{n}\right)\right\rangle\right)=\frac{1}{\kappa},
\nonumber\\
&&\frac{U}{2\left(\beta{N}\right)^{2}}\sum_{{\bf{k}}\nu_{n},\sigma}\left(\left\langle \bar{a}_{1{\bf{k}}\uparrow}\left(\nu_{n}\right)a_{1{\bf{k}}\uparrow}\left(\nu_{n}\right)\right\rangle-\left\langle\bar{a}_{1{\bf{k}}\downarrow}\left(\nu_{n}\right)a_{1{\bf{k}}\downarrow}\left(\nu_{n}\right)\right\rangle\right)=\Delta_{\rm AFM},
\nonumber\\
&&\frac{1}{\left(\beta{N}\right)^{2}}\sum_{{\bf{k}}\nu_{n},\sigma}\left(\left\langle \bar{a}_{1{\bf{k}}\sigma}\left(\nu_{n}\right)a_{1{\bf{k}}\sigma}\left(\nu_{n}\right)\right\rangle-\left\langle \bar{b}_{1{\bf{k}}\sigma}\left(\nu_{n}\right)b_{1{\bf{k}}\sigma}\left(\nu_{n}\right)\right\rangle\right)=\delta{\bar{n}}_{1},
\nonumber\\
&&\frac{1}{\left(\beta{N}\right)^{2}}\sum_{{\bf{k}}\nu_{n},\sigma}\left(\left\langle \bar{a}_{2{\bf{k}}\sigma}\left(\nu_{n}\right)a_{2{\bf{k}}\sigma}\left(\nu_{n}\right)\right\rangle-\left\langle \bar{b}_{2{\bf{k}}\sigma}\left(\nu_{n}\right)b_{2{\bf{k}}\sigma}\left(\nu_{n}\right)\right\rangle\right)=\delta{\bar{n}}_{2},
\nonumber\\
&&\frac{W}{\left(\beta{N}\right)^{2}}\sum_{{\bf{k}}\nu_{n}}\left\langle \bar{a}_{2{\bf{k}}\uparrow}\left(\nu_{n}\right)a_{1{\bf{k}}\uparrow}\left(\nu_{n}\right)\right\rangle+\left\langle \bar{a}_{1{\bf{k}}\uparrow}\left(\nu_{n}\right)a_{2{\bf{k}}\uparrow}\left(\nu_{n}\right)\right\rangle+\left\langle \bar{b}_{2{\bf{k}}\uparrow}\left(\nu_{n}\right)b_{1{\bf{k}}\uparrow}\left(\nu_{n}\right)\right\rangle+\left\langle \bar{b}_{1{\bf{k}}\uparrow}\left(\nu_{n}\right)b_{2{\bf{k}}\uparrow}\left(\nu_{n}\right)\right\rangle=4\Delta_{\uparrow},
\nonumber\\
&&\frac{W}{\left(\beta{N}\right)^{2}}\sum_{{\bf{k}}\nu_{n}}\left\langle \bar{a}_{2{\bf{k}}\downarrow}\left(\nu_{n}\right)a_{1{\bf{k}}\downarrow}\left(\nu_{n}\right)\right\rangle+\left\langle \bar{a}_{1{\bf{k}}\downarrow}\left(\nu_{n}\right)a_{2{\bf{k}}\downarrow}\left(\nu_{n}\right)\right\rangle+\left\langle \bar{b}_{2{\bf{k}}\downarrow}\left(\nu_{n}\right)b_{1{\bf{k}}\downarrow}\left(\nu_{n}\right)\right\rangle+\left\langle \bar{b}_{1{\bf{k}}\uparrow}\left(\nu_{n}\right)b_{2{\bf{k}}\downarrow}\left(\nu_{n}\right)\right\rangle=4\Delta_{\downarrow}, 
\label{Equation_A_3}
\end{eqnarray}
\end{widetext}
where 
\begin{eqnarray}
\Delta_{\uparrow}=\frac{1}{4}\left(\Delta^{\rm A}_{\uparrow}+\bar{\Delta}^{\rm A}_{\uparrow}+\Delta^{\rm B}_{\uparrow}+\bar{\Delta}^{\rm B}_{\uparrow}\right),
\nonumber\\
\Delta_{\downarrow}=\frac{1}{4}\left(\Delta^{\rm A}_{\downarrow}+\bar{\Delta}^{\rm A}_{\downarrow}+\Delta^{\rm B}_{\downarrow}+\bar{\Delta}^{\rm B}_{\downarrow}\right).
\label{Equation_A_4}
\end{eqnarray}

Then, the functional derivatives of the partition function in Eq.(\ref{Equation_A_1}) with respect to the source terms give us the averages entering in the right-sides of the equation Eq.(\ref{Equation_A_3}). We obtain
\begin{eqnarray}
\left\langle \bar{a}_{1{\bf{k}}\sigma}\left(\nu_{n}\right)a_{1{\bf{k}}\sigma}\left(\nu_{n}\right)\right\rangle=2{\cal{D}}_{11{\bf{k}}\sigma}\left(\nu_{n}\right),
\nonumber\\
\left\langle \bar{b}_{1{\bf{k}}\sigma}\left(\nu_{n}\right)b_{1{\bf{k}}\sigma}\left(\nu_{n}\right)\right\rangle=2{\cal{D}}_{22{\bf{k}}\sigma}\left(\nu_{n}\right),
\nonumber\\
\left\langle \bar{a}_{2{\bf{k}}\sigma}\left(\nu_{n}\right)a_{2{\bf{k}}\sigma}\left(\nu_{n}\right)\right\rangle=2{\cal{D}}_{33{\bf{k}}\sigma}\left(\nu_{n}\right),
\nonumber\\
\left\langle \bar{b}_{2{\bf{k}}\sigma}\left(\nu_{n}\right)b_{2{\bf{k}}\sigma}\left(\nu_{n}\right)\right\rangle=2{\cal{D}}_{44{\bf{k}}\sigma}\left(\nu_{n}\right),
\nonumber\\
\left\langle \bar{a}_{2{\bf{k}}\sigma}\left(\nu_{n}\right)a_{1{\bf{k}}\sigma}\left(\nu_{n}\right)\right\rangle=2{\cal{D}}_{13{\bf{k}}\sigma}\left(\nu_{n}\right),
\nonumber\\
\left\langle \bar{a}_{1{\bf{k}}\sigma}\left(\nu_{n}\right)a_{2{\bf{k}}\sigma}\left(\nu_{n}\right)\right\rangle=2{\cal{D}}_{31{\bf{k}}\sigma}\left(\nu_{n}\right),
\nonumber\\
\left\langle \bar{b}_{2{\bf{k}}\sigma}\left(\nu_{n}\right)b_{1{\bf{k}}\sigma}\left(\nu_{n}\right)\right\rangle=2{\cal{D}}_{24{\bf{k}}\sigma}\left(\nu_{n}\right),
\nonumber\\
\left\langle \bar{b}_{1{\bf{k}}\sigma}\left(\nu_{n}\right)b_{2{\bf{k}}\sigma}\left(\nu_{n}\right)\right\rangle=2{\cal{D}}_{42{\bf{k}}\sigma}\left(\nu_{n}\right).
\label{Equation_A_5}
\end{eqnarray}
On the other hand, the explicit calculation of the inverse matrix elements ${\cal{D}}_{11{\bf{k}}\sigma}\left(\nu_{n}\right)$, ${\cal{D}}_{22{\bf{k}}\sigma}\left(\nu_{n}\right)$, ${\cal{D}}_{33{\bf{k}}\sigma}\left(\nu_{n}\right)$, ${\cal{D}}_{44{\bf{k}}\sigma}\left(\nu_{n}\right)$, ${\cal{D}}_{13{\bf{k}}\sigma}\left(\nu_{n}\right)$, ${\cal{D}}_{31{\bf{k}}\sigma}\left(\nu_{n}\right)$, ${\cal{D}}_{24{\bf{k}}\sigma}\left(\nu_{n}\right)$ and ${\cal{D}}_{42{\bf{k}}\sigma}\left(\nu_{n}\right)$ for the spin-$\uparrow$ and spin-$\downarrow$ directions permits to write the system of equations in Eq.(\ref{Equation_A_3}) in a more compact form:
\begin{eqnarray}
&&\frac{1}{\beta{N}}\sum_{{\bf{k}}\nu_{n}}\left[\frac{{\cal{P}}^{\left(3\right)}_{\mu,\uparrow}\left({\bf{k}},i\nu_{n}\right)}{\det{{\cal{G}}^{-1}_{{\bf{k}}\uparrow}\left(\nu_{n}\right)}}+\frac{{\cal{P}}^{\left(3\right)}_{\mu,\downarrow}\left({\bf{k}},i\nu_{n}\right)}{\det{{\cal{G}}^{-1}_{{\bf{k}}\downarrow}\left(\nu_{n}\right)}}\right]=\frac{2}{\kappa},
\nonumber\\
&&\frac{1}{\beta{N}}\sum_{{\bf{k}}\nu_{n}}\left[\frac{{\cal{P}}^{\left(2\right)}_{\delta{\bar{n}}_{1},\uparrow}\left({\bf{k}},i\nu_{n}\right)}{\det{{\cal{G}}^{-1}_{{\bf{k}}\uparrow}\left(\nu_{n}\right)}}+\frac{{\cal{P}}^{\left(2\right)}_{\delta{\bar{n}}_{1},\downarrow}\left({\bf{k}},i\nu_{n}\right)}{\det{{\cal{G}}^{-1}_{{\bf{k}}\downarrow}\left(\nu_{n}\right)}}\right]=\delta{\bar{n}}_{1},
\nonumber\\
&&\frac{1}{\beta{N}}\sum_{{\bf{k}}\nu_{n}}\left[\frac{{\cal{P}}^{\left(2\right)}_{\delta{\bar{n}}_{2},\uparrow}\left({\bf{k}},i\nu_{n}\right)}{\det{{\cal{G}}^{-1}_{{\bf{k}}\uparrow}\left(\nu_{n}\right)}}+\frac{{\cal{P}}^{\left(2\right)}_{\delta{\bar{n}}_{2},\downarrow}\left({\bf{k}},i\nu_{n}\right)}{\det{{\cal{G}}^{-1}_{{\bf{k}}\downarrow}\left(\nu_{n}\right)}}\right]=\delta{\bar{n}}_{2},
\nonumber\\
&&\frac{U}{\beta{N}}\sum_{{\bf{k}}\nu_{n}}\left[\frac{{\cal{P}}^{\left(3\right)}_{\Delta_{\rm AFM},\uparrow}\left({\bf{k}},i\nu_{n}\right)}{\det{{\cal{G}}^{-1}_{{\bf{k}}\uparrow}\left(\nu_{n}\right)}}-\frac{{\cal{P}}^{\left(3\right)}_{\Delta_{\rm AFM},\downarrow}\left({\bf{k}},i\nu_{n}\right)}{\det{{\cal{G}}^{-1}_{{\bf{k}}\downarrow}\left(\nu_{n}\right)}}\right]=\Delta_{\rm AFM},
\nonumber\\
&&\frac{W}{4\beta{N}}\sum_{{\bf{k}}\nu_{n}}\frac{{\cal{P}}^{\left(2\right)}_{\Delta_{\uparrow}}\left({\bf{k}},i\nu_{n}\right)}{\det{{\cal{G}}^{-1}_{{\bf{k}}\uparrow}\left(\nu_{n}\right)}}=\Delta_{\uparrow},
\nonumber\\
&&\frac{W}{4\beta{N}}\sum_{{\bf{k}}\nu_{n}}\frac{{\cal{P}}^{\left(2\right)}_{\Delta_{\downarrow}}\left({\bf{k}},i\nu_{n}\right)}{\det{{\cal{G}}^{-1}_{{\bf{k}}\downarrow}\left(\nu_{n}\right)}}=\Delta_{\downarrow}.
\nonumber\\
\label{Equation_A_6}
\end{eqnarray}
We have defined a series of polynomials in Eq.(\ref{Equation_A_6}):
\begin{widetext}
\begin{eqnarray}
&&{\cal{P}}^{\left(3\right)}_{\mu,\uparrow}\left({\bf{k}},i\nu_{n}\right)=4x^{3}-12\mu_{0\uparrow}x^{2}-2\left(2|\Delta_{\uparrow}+\gamma_1|^{2}+2|\tilde{\gamma}_{{\bf{k}}}|^{2}+\Delta^{2}_{1}+\Delta^{2}_{2}-6\mu^{2}_{0\uparrow}\right)x
\nonumber\\
&&+2\mu_{0\uparrow}\left(2|\Delta_{\uparrow}+\gamma_1|^{2}+2|\tilde{\gamma}_{{\bf{k}}}|^{2}+\Delta^{2}_{1}+\Delta^{2}_{2}-2\mu^{2}_{0\uparrow}\right),
\nonumber\\
&&{\cal{P}}^{\left(3\right)}_{\mu,\downarrow}\left({\bf{k}},i\nu_{n}\right)=4x^{3}-12\mu_{0\downarrow}x^{2}-2\left(2|\Delta_{\downarrow}+\gamma_1|^{2}+2|\tilde{\gamma}_{{\bf{k}}}|^{2}+\Delta^{2}_{3}+\Delta^{2}_{4}-6\mu^{2}_{0\downarrow}\right)x
\nonumber\\
&&+2\mu_{0\downarrow}\left(2|\Delta_{\downarrow}+\gamma_1|^{2}+2|\tilde{\gamma}_{{\bf{k}}}|^{2}+\Delta^{2}_{3}+\Delta^{2}_{4}-2\mu^{2}_{0\downarrow}\right),
\nonumber\\
&&{\cal{P}}^{\left(2\right)}_{\delta{\bar{n}}_{1},\uparrow}\left({\bf{k}},i\nu_{n}\right)=2x^{2}\Delta_{1}-4\Delta_{1}\mu_{0\uparrow}x-2\left[|\Delta_{\uparrow}+\gamma_1|^{2}\Delta_{2}+\Delta_{1}\left(|\tilde{\gamma}_{\bf{k}}|^{2}+\Delta^{2}_{2}-\mu^{2}_{0\uparrow}\right)\right],
\nonumber\\
&&{\cal{P}}^{\left(2\right)}_{\delta{\bar{n}}_{1},\downarrow}\left({\bf{k}},i\nu_{n}\right)=-2x^{2}\Delta_{3}+4\Delta_{3}\mu_{0\downarrow}x+2\left[|\Delta_{\downarrow}+\gamma_1|^{2}\Delta_{4}+\Delta_{3}\left(|\tilde{\gamma}_{\bf{k}}|^{2}+\Delta^{2}_{4}-\mu^{2}_{0\downarrow}\right)\right],
\nonumber\\
&&{\cal{P}}^{\left(2\right)}_{\delta{\bar{n}}_{2},\uparrow}\left({\bf{k}},i\nu_{n}\right)=-2x^{2}\Delta_{2}+4\Delta_{2}\mu_{0\uparrow}x+2\left[|\Delta_{\uparrow}+\gamma_1|^{2}\Delta_{1}+\Delta_{2}\left(|\tilde{\gamma}_{\bf{k}}|^{2}+\Delta^{2}_{1}-\mu^{2}_{0\uparrow}\right)\right],
\nonumber\\
&&{\cal{P}}^{\left(2\right)}_{\delta{\bar{n}}_{2},\downarrow}\left({\bf{k}},i\nu_{n}\right)=2x^{2}\Delta_{4}-4\Delta_{4}\mu_{0\downarrow}x-2\left[|\Delta_{\downarrow}+\gamma_1|^{2}\Delta_{3}+\Delta_{4}\left(|\tilde{\gamma}_{\bf{k}}|^{2}+\Delta^{2}_{3}-\mu^{2}_{0\downarrow}\right)\right],
\nonumber\\
&&{\cal{P}}^{\left(3\right)}_{\Delta_{\rm AFM},\uparrow}\left({\bf{k}},i\nu_{n}\right)=x^{3}+\left(\Delta_{1}-3\mu_{0\uparrow}\right)x^{2}+\left(3\mu^{2}_{0\uparrow}-|\tilde{\gamma}_{{\bf{k}}}|^{2}-\Delta^{2}_{2}-2\Delta_{1}\mu_{0\uparrow}-|\Delta_{\uparrow}+\gamma_1|^{2}\right)x
\nonumber\\
&&+\left(\mu_{0\uparrow}-\Delta_{2}\right)|\Delta_{\uparrow}+\gamma_1|^{2}+\left(\Delta_{1}-\mu_{0\uparrow}\right)\left(\mu^{2}_{0\uparrow}-\Delta^{2}_{2}-|\tilde{\gamma}_{{\bf{k}}}|^{2}\right),
\nonumber\\
&&{\cal{P}}^{\left(3\right)}_{\Delta_{\rm AFM},\downarrow}\left({\bf{k}},i\nu_{n}\right)=x^{3}+\left(-\Delta_{3}-3\mu_{0\downarrow}\right)x^{2}+\left(3\mu^{2}_{0\downarrow}-|\tilde{\gamma}_{{\bf{k}}}|^{2}-\Delta^{2}_{4}+2\Delta_{3}\mu_{0\downarrow}-|\Delta_{\downarrow}+\gamma_1|^{2}\right)x
\nonumber\\
&&+\left(\mu_{0\downarrow}+\Delta_{4}\right)|\Delta_{\downarrow}+\gamma_1|^{2}+\left(\Delta_{3}-\mu_{0\downarrow}\right)\left(-\mu^{2}_{0\downarrow}+\Delta^{2}_{4}+|\tilde{\gamma}_{{\bf{k}}}|^{2}\right),
\nonumber\\
&&{\cal{P}}^{\left(2\right)}_{\Delta{\uparrow}}\left({\bf{k}},i\nu_{n}\right)=4\Delta_{\uparrow}x^{2}-8\Delta_{\uparrow}\mu_{0\uparrow}x+4\Delta_{\uparrow}\left(|\tilde{\gamma}_{{\bf{k}}}|^{2}-\Delta_{1}\Delta_{2}-|\Delta_{\uparrow}+\gamma_1|^{2}+\mu^{2}_{0\uparrow}\right),
\nonumber\\
&&{\cal{P}}^{\left(2\right)}_{\Delta{\downarrow}}\left({\bf{k}},i\nu_{n}\right)=4\Delta_{\downarrow}x^{2}-8\Delta_{\downarrow}\mu_{0\downarrow}x+4\Delta_{\downarrow}\left(|\tilde{\gamma}_{{\bf{k}}}|^{2}-\Delta_{3}\Delta_{4}-|\Delta_{\downarrow}+\gamma_1|^{2}+\mu^{2}_{0\downarrow}\right),
\label{Equation_A_7}
\end{eqnarray}
\end{widetext}
where $x=-i\nu_{n}$.

Furthermore, we put the expressions in Eq.(\ref{Equation_A_7}) into Eq.(\ref{Equation_A_6}) and we evaluate the fractions in the left-hand side in Eq.(\ref{Equation_A_6}). After this, we perform the summations over the fermionic Matsubara frequencies $\nu_{n}$, via the following formula
\begin{eqnarray}
k_{B}T\sum_{\nu_{n}}\frac{1}{i\nu_{n}-x}=-n_{\rm F}\left(x\right),
\label{Equation_A_8}
\end{eqnarray}
where $n_{\rm F}\left(x\right)$ is Fermi-Dirac distribution function defined in Eq.(\ref{Equation_37}).
We obtain the system of self-consistent equations in the form, given in Eq.(\ref{Equation_36}), in the Section \ref{sec:Section_2_3} and for the coefficients $\alpha^{\sigma}_{i{\bf{k}}}$, $\beta^{\sigma}_{ji{\bf{k}}}$, $\gamma^{\sigma}_{i{\bf{k}}}$, and $\delta^{\sigma}_{i{\bf{k}}}$ (where $j=1,2$ and $i=1,...,4$) we get 
\begin{widetext}
\begin{eqnarray}
\footnotesize
\arraycolsep=0pt
\medmuskip = 0mu
\alpha^{\sigma}_{i{\bf{k}}}
=\left\{
\begin{array}{cc}
\displaystyle  & \frac{\left(-1\right)^{i+1}}{\varepsilon_{1\sigma}({\bf{k}})-\varepsilon_{2\sigma}({\bf{k}})}\prod_{j=3,4}\frac{{\cal{P}}^{\left(3\right)}_{\mu,\uparrow}\left({\bf{k}},\varepsilon_{i\sigma}({\bf{k}})\right)}{\varepsilon_{i\sigma}({\bf{k}})-\varepsilon_{j\sigma}({\bf{k}})}, \ \ \ $if$ \ \ \  i=1,2.
\newline\\
\newline\\
\displaystyle  & \frac{\left(-1\right)^{i+1}}{\varepsilon_{3\sigma}({\bf{k}})-\varepsilon_{4\sigma}({\bf{k}})}\prod_{j=1,2}\frac{{\cal{P}}^{\left(3\right)}_{\mu,\uparrow}\left({\bf{k}},\varepsilon_{i\sigma}({\bf{k}})\right)}{\varepsilon_{i\sigma}({\bf{k}})-\varepsilon_{j\sigma}({\bf{k}})}, \ \ \ $if$ \ \ \  i=3,4,
\end{array}\right.
\label{Equation_A_9}
\end{eqnarray}
\begin{eqnarray}
\footnotesize
\arraycolsep=0pt
\medmuskip = 0mu
\beta^{\sigma}_{1i{\bf{k}}}
=\left\{
\begin{array}{cc}
\displaystyle  & \frac{\left(-1\right)^{i+1}}{\varepsilon_{1\sigma}({\bf{k}})-\varepsilon_{2\sigma}({\bf{k}})}\prod_{j=3,4}\frac{{\cal{P}}^{\left(2\right)}_{\delta{\bar{n}}_{1},\sigma}\left({\bf{k}},\varepsilon_{i\sigma}({\bf{k}})\right)}{\varepsilon_{i\sigma}({\bf{k}})-\varepsilon_{j\sigma}({\bf{k}})}, \ \ \ $if$ \ \ \  i=1,2.
\newline\\
\newline\\
\displaystyle  & \frac{\left(-1\right)^{i+1}}{\varepsilon_{3\sigma}({\bf{k}})-\varepsilon_{4\sigma}({\bf{k}})}\prod_{j=1,2}\frac{{\cal{P}}^{\left(2\right)}_{\delta{\bar{n}}_{1},\sigma}\left({\bf{k}},\varepsilon_{i\sigma}({\bf{k}})\right)}{\varepsilon_{i\sigma}({\bf{k}})-\varepsilon_{j\sigma}({\bf{k}})}, \ \ \ $if$ \ \ \  i=3,4,
\end{array}\right.
\label{Equation_A_10}
\end{eqnarray}
\begin{eqnarray}
\footnotesize
\arraycolsep=0pt
\medmuskip = 0mu
\beta^{\sigma}_{2i{\bf{k}}}
=\left\{
\begin{array}{cc}
\displaystyle  & \frac{\left(-1\right)^{i+1}}{\varepsilon_{1\sigma}({\bf{k}})-\varepsilon_{2\sigma}({\bf{k}})}\prod_{j=3,4}\frac{{\cal{P}}^{\left(2\right)}_{\delta{\bar{n}}_{2},\sigma}\left({\bf{k}},\varepsilon_{i\sigma}({\bf{k}})\right)}{\varepsilon_{i\sigma}({\bf{k}})-\varepsilon_{j\sigma}({\bf{k}})}, \ \ \ $if$ \ \ \  i=1,2.
\newline\\
\newline\\
\displaystyle  & \frac{\left(-1\right)^{i+1}}{\varepsilon_{3\sigma}({\bf{k}})-\varepsilon_{4\sigma}({\bf{k}})}\prod_{j=1,2}\frac{{\cal{P}}^{\left(2\right)}_{\delta{\bar{n}}_{2},\sigma}\left({\bf{k}},\varepsilon_{i\sigma}({\bf{k}})\right)}{\varepsilon_{i\sigma}({\bf{k}})-\varepsilon_{j\sigma}({\bf{k}})}, \ \ \ $if$ \ \ \  i=3,4,
\end{array}\right.
\label{Equation_A_11}
\end{eqnarray}
\begin{eqnarray}
\footnotesize
\arraycolsep=0pt
\medmuskip = 0mu
\gamma^{\sigma}_{i{\bf{k}}}
=\left\{
\begin{array}{cc}
\displaystyle  & \frac{\left(-1\right)^{i+1}}{\varepsilon_{1\sigma}({\bf{k}})-\varepsilon_{2\sigma}({\bf{k}})}\prod_{j=3,4}\frac{{\cal{P}}^{\left(3\right)}_{\Delta_{\rm AFM},\sigma}\left({\bf{k}},\varepsilon_{i\sigma}({\bf{k}})\right)}{\varepsilon_{i\sigma}({\bf{k}})-\varepsilon_{j\sigma}({\bf{k}})}, \ \ \ $if$ \ \ \  i=1,2.
\newline\\
\newline\\
\displaystyle  & \frac{\left(-1\right)^{i+1}}{\varepsilon_{3\sigma}({\bf{k}})-\varepsilon_{4\sigma}({\bf{k}})}\prod_{j=1,2}\frac{{\cal{P}}^{\left(3\right)}_{\Delta_{\rm AFM},\sigma}\left({\bf{k}},\varepsilon_{i\sigma}({\bf{k}})\right)}{\varepsilon_{i\sigma}({\bf{k}})-\varepsilon_{j\sigma}({\bf{k}})}, \ \ \ $if$ \ \ \  i=3,4,
\end{array}\right.
\label{Equation_A_12}
\end{eqnarray}
\begin{eqnarray}
\footnotesize
\arraycolsep=0pt
\medmuskip = 0mu
\delta^{\sigma}_{i{\bf{k}}}
=\left\{
\begin{array}{cc}
\displaystyle  & \frac{\left(-1\right)^{i+1}}{\varepsilon_{1\sigma}({\bf{k}})-\varepsilon_{2\sigma}({\bf{k}})}\prod_{j=3,4}\frac{{\cal{P}}^{\left(2\right)}_{\Delta{\uparrow}}\left({\bf{k}},\varepsilon_{i\sigma}({\bf{k}})\right)}{\varepsilon_{i\sigma}({\bf{k}})-\varepsilon_{j\sigma}({\bf{k}})}, \ \ \ $if$ \ \ \  i=1,2.
\newline\\
\newline\\
\displaystyle  & \frac{\left(-1\right)^{i+1}}{\varepsilon_{3\sigma}({\bf{k}})-\varepsilon_{4\sigma}({\bf{k}})}\prod_{j=1,2}\frac{{\cal{P}}^{\left(2\right)}_{\Delta{\downarrow}}\left({\bf{k}},\varepsilon_{i\sigma}({\bf{k}})\right)}{\varepsilon_{i\sigma}({\bf{k}})-\varepsilon_{j\sigma}({\bf{k}})}, \ \ \ $if$ \ \ \  i=3,4.
\end{array}\right.
\label{Equation_A_13}
\end{eqnarray}
\end{widetext}
\bibliography{references_authors}

\end{document}